\documentclass[12pt]{article}

\usepackage{epsfig}
\usepackage{cite}
\usepackage{amsmath, amssymb, amsfonts}
\usepackage{color}
\usepackage{latexsym}  
\usepackage{graphicx}
\usepackage{cancel}
\usepackage[colorlinks,bookmarks]{hyperref}
\hypersetup{pdfpagemode=UseNone, pdfstartview=FitH, linkcolor=blue, citecolor=red, urlcolor=blue}

\def\be{\begin{equation}}
\def\ee{\end{equation}}
\def\ba{\begin{eqnarray}}
\def\ea{\end{eqnarray}}

\def\bdm{\begin{displaymath}}
\def\edm{\end{displaymath}}

\def\bq{\begin{quote}}
\def\eq{\end{quote}}

 at 10truept

\newcommand{\Mpl}{M_{\mathrm{Pl}}}

\newcommand{\bea}{\begin{eqnarray}}
\newcommand{\eea}{\end{eqnarray}}

\newcommand{\bi}{\begin{itemize}}
\newcommand{\ei}{\end{itemize}}

\newcommand{\beq}{\begin{equation}}
\newcommand{\eeq}{\end{equation}}
\newcommand{\beqa}{\begin{eqnarray}}
\newcommand{\eeqa}{\end{eqnarray}}
\newcommand{\mpl}{\Mpl}

\newcommand\unity{1\!\!1}

\def\12{{1 \over 2}}

\def\ltap{\ \raise.3ex\hbox{$<$\kern-.75em\lower1ex\hbox{$\sim$}}\ }
\def\gtap{\ \raise.3ex\hbox{$>$\kern-.75em\lower1ex\hbox{$\sim$}}\ }
\def\gl{\ \raise.5ex\hbox{$>$}\kern-.8em\lower.5ex\hbox{$<$}\ }
\def\roughly#1{\raise.3ex\hbox{$#1$\kern-.75em\lower1ex\hbox{$\sim$}}}

\begin{document}

\thispagestyle{empty}
\begin{flushright}
May 2026
\end{flushright}
\vspace*{1.25cm}
\begin{center}

{\Large \bf CMB Birefringence from Vacuum Interfaces}

\vspace*{1.35cm} {\large
Nemanja Kaloper\footnote{\tt
kaloper@physics.ucdavis.edu} }\\
\vspace{.5cm}
{\em QMAP, Department of Physics and Astronomy, University of
California}\\
\vspace{.05cm}
{\em Davis, CA 95616, USA}\\

\vspace{1.25cm} ABSTRACT
\end{center}
Hints of cosmic microwave background polarization rotation
($\Delta\vartheta \sim 10^{-3}$ rad) are commonly attributed to late-time
dynamics of ultralight axions. We show that such ultralight degrees of freedom
are not required. Polarization rotation naturally arises as a
geometric interface phase acquired when photons cross
interfaces between topologically distinct dark sector vacua. The effect is a
discrete phase shift fixed by the normalization of a wall-supported
electromagnetic Chern--Simons interaction and protected by an emergent
$1$-form symmetry of the low energy effective theory. 
This mechanism reproduces the familiar adiabatic rotation induced by light
axion domain walls, but persists for arbitrarily thin walls where the axion is
heavy or absent. In this regime the rotation manifests as a Pancharatnam phase
localized at vacuum interfaces, independent of redshift and photon frequency
below a natural ultraviolet cutoff. Cosmic birefringence thus emerges as a
probe of vacuum structure in the dark sector, rather
than of light-field dynamics.

\vfill \setcounter{page}{0} \setcounter{footnote}{0}

\vspace{1cm}

\newpage

\section{Introduction} 

Tantalizing observational hints \cite{Komatsu:2022nvu}
of cosmic microwave background (CMB) polarization rotation,
at the level of $\Delta\vartheta \sim 10^{-3}\,\mathrm{rad}$, have attracted
increasing attention in recent years. Standard lore attributes such signals to
late-time dynamics of ultralight fields, most notably axion-like pseudoscalars
\cite{Huang:1985tt,Harari:1992ea} (see also 
\cite{Carroll:1998zi,Lue:1998mq,Pospelov:2008gg,Minami:2020odp,Takahashi:2020tqv,Diego-Palazuelos:2022dsq,Ferreira:2023jbu}),
which couple to the CP-odd electromagnetic operator
$\epsilon_{\mu\nu\lambda\sigma} F^{\mu\nu} F^{\lambda\sigma}$.
In many scenarios an axion evolving in time axion couples differently to the two
electromagnetic helicities, inducing a gradual and cumulative rotation of the
polarization plane as light propagates across cosmological distances. The effect
is intrinsically adiabatic and depends sensitively on the spacetime profile and
time evolution of the scalar field. As a consequence, producing an observable
rotation requires the axion to be extremely light, with mass roughly below the Hubble
scale at last scattering,
$m \lesssim H_{\tt LSS} \sim 10^{-28}\,\mathrm{eV}$.

This constraint can be seen directly. Since the last scattering surface occurs
during matter domination, CMB photons propagate predominantly through a
matter-dominated universe. Truncating the axion potential to its mass term, an
initially displaced axion obeys
\be
\ddot\phi + 3H\dot\phi + m^2\phi \simeq 0 \, .
\ee
In matter domination, when the axion is a spectator, this equation admits the exact solution
\be
\phi(t) = {\tt a}\,f_\phi t_i \frac{\cos[m(t-t_i)]}{t} \, , 
\ee
where ${\tt a}$ is a ${\cal O}(1)$ numbers which 
parameterize initial conditions, $t_i$ the time $\phi$ begins to oscillate 
and $f_\phi$ is the axion decay constant.
The total polarization rotation induced by such an evolution is bounded by
\be
\Delta\vartheta \lesssim {\tt a} mt_i \int_{t_i}^{t_f} \!dt\,\frac{\sin[m(t-t_i)]}{t} \, .
\ee
Over the relevant interval $\Delta t = t_f - t_i \lesssim H^{-1}$, the factor
$1/t$ decreases slowly, while $\cos(mt)$ oscillates rapidly when $m\gg H$.
By the Riemann--Lebesgue lemma the contributions to the integral will be suppressed except inside
an interval of a single period around the initial time,
yielding at most $\Delta\vartheta \simeq {\tt a} mt_i/mt_i \simeq {\tt a}$. 
If $m> H_{\tt LSS}$ this will occur before the CMB is around, and even if $\phi$ is still
oscillating after the last scattering surface, the imprint on the CMB will be too suppressed by premature 
release of the axion $\phi$.
Thus we need $m \lesssim H$, where $H \lesssim H_{\tt LSS}$. 
Up to ${\cal O}(1)$ numerical factors, this requires
$m \lesssim 10^{-28}\,\mathrm{eV}$ for an observable effect.
Moreover, any such signal depends sensitively on the detailed history of the
axion evolution and can be contaminated by inhomogeneities in geometry, matter
distribution, or local dynamics. In practice, producing polarization rotation
from time-evolving fields demands the presence of ultralight scalars\footnote{Note, that these 
solutions oscillate around a single broad axion minimum. They can be thought of as the configurations 
which didn't have enough energy to extrapolate to the next minimum, and build a wall.}.

In this work we show that ultralight degrees of freedom are not required to generate observable
cosmic birefringence. Instead, cosmic birefringence mechanisms fall into two distinct
universality classes. In the first, birefringence is dynamical and controlled by broken
$0$-form symmetries, such as the axion shift symmetry, requiring light axions and adiabatic
evolution. In the second, birefringence arises from vacuum interfaces and is controlled by
emergent $1$-form symmetries of the low-energy effective theory. The mechanism presented here
provides a unified description of both cases and an explicit realization of the interface-driven
class. The relevant symmetry structure arises only in the infrared effective theory; in the
ultraviolet the topology of the full Hilbert space need not be directly accessible to
low-energy probes.

Our starting point is a reinterpretation of axion-induced polarization rotation at low
frequencies as a symmetry-protected geometric interface phase associated with photon
propagation through different $\theta$-vacua, rather than as a purely dynamical effect driven by
time evolution of an ultralight pseudoscalar. Once cast in this form, the
mechanism naturally generalizes beyond slowly varying scalar backgrounds, relaxing the
requirement that birefringence be adiabatic. In particular, when ultralight axions are absent,
polarization rotation arises as a phase accrued when photons traverse
interfaces between topologically distinct vacuum regions. Crossing such an interface induces
a discrete twist of the polarization plane, as originally 
discovered by Shivaramakrishnan Pancharatnam in
optically active materials in 1956 \cite{Pancharatnam:1956url,Pancharatnam:1956url2}.
If we are surrounded by one such wall, or an array of many 
connected ones, the effect covers the whole sky. 

An additional attractive feature of this mechanism is that the magnitude of the 
effect does not rely on ultralight mass scales. The polarization rotation is 
controlled by the coefficient of the wall-supported electromagnetic Chern--Simons interaction, 
which can naturally inherit anomaly-induced suppression factors of 
order $\alpha_{\rm QED}/2\pi \sim 10^{-3}$ from heavy charged matter 
integrated out of the low-energy theory \cite{Kaloper:2025goq,Kaloper:2026slg,Kaloper:2026gib}. 
Since this coupling arises from anomaly-induced interactions generated by heavy charged states, 
its coefficient is radiatively stable and protected by the Adler--Bardeen nonrenormalization theorem 
\cite{Adler:1969er}. Remarkably, the observed hint of birefringence 
at the level of a few times $10^{-3}$ radians is naturally of the expected 
magnitude in this framework.

After reviewing the adiabatic regime in which light axions are present, we consider the axion
decoupling limit. We focus on the simplest case of a single domain-wall crossing that induces
a discrete polarization rotation. This setting arises naturally in the context of Discretely
Evanescent Dark Energy \cite{Kaloper:2025goq,Kaloper:2026slg,Kaloper:2026gib}. More general
vacuum structures and multiple crossings are straightforward extensions and are discussed in
section 6 . We develop a unified quantum-mechanical description of probe photons, such as
the CMB, propagating through backgrounds with a varying vacuum angle $\theta$ along the line
of sight, irrespective of whether the variation is smooth and adiabatic or discontinuous due
to the discharge of a top-form flux. This approach simplifies the derivation of polarization
rotation \cite{Huang:1985tt,Harari:1992ea,Ganoulis:1986rd,Favitta:2023hlx,Agrawal:2023sbp,Blasi:2024xvj} and
makes manifest that the only essential criterion is whether a photon crosses from one vacuum
region to another. Photons propagating entirely within a single vacuum experience no rotation.

The resulting phase change could be interpreted as a holonomy in polarization
space arising from interface-induced matching of electromagnetic polarization states
across an interface separating topologically distinct dark-sector vacua. The structure of the
interface is constrained by an emergent $1$-form symmetry \cite{Gaiotto:2014kfa}, which
protects the allowed phase jump and renders it insensitive to local details of the
interpolation between vacua \cite{Hidaka:2020iaz,Brennan:2020ehu}.
Consequently, in the absence of adiabatically varying light axions, polarization rotation
occurs only at localized vacuum interfaces, and its magnitude is determined by global
topological data rather than by local dynamics or the cosmological history between crossings.

An important question is whether vacuum interfaces capable of producing
observable birefringence can arise consistently in cosmology. In this paper
we focus on the class of low-tension, late-time domain walls arising in
Discretely Evanescent Dark Energy
\cite{Kaloper:2025goq,Kaloper:2026slg,Kaloper:2026gib}. These walls are
characterized by scales in the range
${\cal M}\sim10^{-3}\,{\rm eV}$--${\rm eV}$ and naturally evade the
standard cosmological difficulties associated with high-tension domain-wall
networks. The same scale yields vacuum-transition rates of order
$\Gamma\sim H_0^4$, making such transitions cosmologically relevant near
the present epoch, and plays the role of the cutoff of the low energy EFT. 
Hence such walls are naturally produced in a universe whose age is comparable 
to ours. Detailed analyses of membrane nucleation, vacuum decay, wall evolution,
abundance estimates, and cosmological constraints have been presented in
\cite{Kaloper:2025goq,Kaloper:2026slg,Kaloper:2026gib} and will not be
repeated here. The purpose of the present work is instead to determine the
polarization signal generated when photons cross such interfaces. 

A key feature of this mechanism is the presence of a natural ultraviolet cutoff.
Because the polarization rotation is associated with finite-scale interfaces, sufficiently
high-frequency photons resolve the microscopic wall structure generated by massive degrees of
freedom with anomalous couplings. Above the cutoff, these modes are integrated back into the
theory, screening the effective Chern--Simons interaction and suppressing optical activity
\cite{Kaloper:2025goq,Kaloper:2026slg,Kaloper:2026gib}. In this regime, photon propagation
through the wall becomes effectively sudden \cite{messiah}, eliminating coherent phase
accumulation. The decoupling thus occurs precisely when the effective description breaks down.

Taken together, these results show that cosmic birefringence can arise from the topological
structure of the dark-sector vacuum itself. In this picture, polarization rotation is a
symmetry-protected geometric phase associated with interfaces between distinct vacua,
rather than a dynamical signal of a very light field evolution.

\section{The Low Energy CP-odd Theory and its Vacua}

The purpose of this section is to define the low-energy effective field theory
and its discrete vacuum structure, which together determine the existence and
properties of domain walls relevant for photon propagation and polarization
rotation in subsequent sections.
We consider a gauge-invariant and Lorentz-invariant low-energy effective field
theory, working in the decoupling limit of gravity,
$\mpl \to \infty$, and neglecting spacetime curvature. The action is
\cite{Kaloper:2025goq,Kaloper:2025wgn,Kaloper:2025upu}
\ba 
S &\ni& \int d^4 x \Bigl\{-\frac{1}{4} F_{\mu\nu}^2 - A_\mu J^\mu
- \frac{\zeta}{4! {\cal M}^2} \bigl(\sqrt{\cal X}\frac{\phi}{f_\phi} + {\cal H} \bigr)
\epsilon^{\mu\nu\lambda\sigma} F_{\mu\nu} F_{\lambda\sigma} \nonumber \\
&& \hspace{1.5cm}
+ \frac12 (\partial \phi)^2
- V_{\tt I}\!\left(\frac{\phi}{f_\phi} + \frac{\hat \theta}{2\pi}\right)
- \frac{1}{2} \bigl(\sqrt{\cal X}\frac{\phi}{f_\phi} + {\cal H} \bigr)^2
+ \frac{1}{6} \epsilon^{\mu\nu\lambda\sigma}
\partial_\mu {\cal H}\, {\cal B}_{\nu\lambda\sigma}
\Bigr\} \nonumber \\
&& \hspace{1.5cm}
- {\cal T} \int d^3 \xi \sqrt{ \bigl| \det \left(\eta_{\mu\nu}
\frac{\partial x^\mu}{\partial \xi^a}
\frac{\partial x^\nu}{\partial \xi^b} \right) \bigr| }
- \frac{\cal Q}{6} \int d^3 \xi \,
{\cal B}_{\mu\nu\lambda}
\frac{\partial x^\mu}{\partial \xi^a}
\frac{\partial x^\nu}{\partial \xi^b}
\frac{\partial x^\lambda}{\partial \xi^c} \epsilon^{abc} \, .
\label{cantra}
\ea
Here $A_\mu$ and $F_{\mu\nu}=\partial_\mu A_\nu-\partial_\nu A_\mu$ are the
electromagnetic gauge field and field strength, and $J^\mu$ is the conserved
matter current. For the purposes of this work we set $J^\mu = 0$. Models with the same structure 
have been studied as realization of dark energy and inflation, when coupling to gravity is
restored \cite{Kaloper:2008qs,Kaloper:2008fb,Kaloper:2011jz}.

The field ${\cal H}$ is a pseudoscalar magnetic dual of the electric four-form
field strength ${\cal G}_{\mu\nu\lambda\sigma}
=4\partial_{[\mu}{\cal B}_{\nu\lambda\sigma]}$, with ${\cal B}_{\nu\lambda\sigma}$
the corresponding three-form gauge potential. The higher-form sector is sourced
by membranes of tension ${\cal T}\ge0$ and charge ${\cal Q}$. The scale ${\cal M}$
characterizes the ultraviolet completion of the top-form sector and corresponds
to the strong-coupling scale of the dark gauge theory. The topological
susceptibility ${\cal X}$ generated by strong dynamics fixes
\be
{\cal M} = {\cal X}^{1/4} \, .
\ee
The dimensionless parameter $\zeta$ controls the ratio ${\cal Q}/{\cal M}^2$
\cite{Luscher:1978rn,Gabadadze:1997kj,Gabadadze:2002ff}.

The pseudoscalar $\phi$ plays the role of an axion-like field that mixes with the
dark top-form sector through monodromy induced by strong coupling effects. This
motivates the normalization of the $\phi$--${\cal H}$ interaction by ${\cal X}$.
We allow $\phi$ to couple to additional ultraviolet physics, such as heavy axions
or particles carrying dark microcharges, so that its instanton-induced potential
$V_{\tt I}(\phi/f_\phi+\hat\theta/2\pi)$ need not be aligned with the dark sector.
In this sense $\phi$ is a ``low-quality'' axion: its vacuum structure and mass
need not be controlled by the dark gauge theory alone, and we treat the axion mass
arising from $V_{\tt I}$ as an adjustable parameter.

Equation~\eqref{cantra} should be regarded strictly as an effective field theory
valid below the dark-sector strong-coupling scale ${\cal M}$. A concrete UV origin
is provided by a non-Abelian gauge theory confining at very low energies
\cite{Kaloper:2025goq}. As shown by L\"uscher \cite{Luscher:1978rn}, such theories
naturally generate topological four-forms, whose longitudinal modes correspond to
pseudoscalars when the four-form is massive
\cite{Dvali:2005an,Dvali:2005zk}. Even when all dark propagating degrees of freedom
are integrated out, CP-odd couplings between electromagnetism and the vacuum
structure persist as topological remnants.

Because the electromagnetic Chern--Simons interaction arises from integrating out
heavy states, the effective coupling $\zeta$ is generically scale dependent $\zeta \rightarrow \zeta(k)$.
A convenient way to parameterize this physics is through a wall-localized
ultraviolet completion. For example, we may introduce a heavy phase field $\Phi$
with mass ${\cal M}$ confined to the domain wall, described schematically by the
effective action
\be
S_{\rm wall} \ni
\int d^4x\,\delta({\cal M}z)\!
\left[
\frac12(\partial\Phi)^2
-\frac12{\cal M}^2\Phi^2
+ \tilde g\,{\cal M}^2 f_\phi\,\Phi
- \frac{\hat g}{4!}\frac{\Phi}{f_\Phi}
\epsilon^{\mu\nu\lambda\sigma}F_{\mu\nu}F_{\lambda\sigma}
\right].
\label{phase1}
\ee
At momenta $k \ll {\cal M}$ the heavy mode $\Phi$ can be integrated out, yielding
a wall-localized Chern--Simons interaction with an effective coefficient
$\zeta(k)$ that approaches a constant for $k\!\ll\!{\cal M}$ and is suppressed
for $k\!\gtrsim\!{\cal M}$. Explicitly, solving the $\Phi$ equation of motion in
momentum space shows that $\zeta(k)$ decreases parametrically as
\be
\zeta(k) \simeq \frac{\zeta(0)}{1+k^2/{\cal M}^2} \, , 
\label{zetaeq}
\ee
rendering the wall transparent to sufficiently
high-frequency photons. While the detailed form of $\zeta(k)$ is
model-dependent, this qualitative behavior is robust
\cite{vonDossow:2025bwr}. On the other hand, 
below the cutoff, where $\zeta(k)\simeq\zeta(0)$, localization of the interaction
to the wall introduces an extra power of momentum in the numerator of the photon
scattering kernel. This feature will play a central role in the emergence of a
frequency-independent polarization rotation below the cutoff ${\cal M}$, and ultimately in the
interpretation of the effect as a symmetry-protected geometric interface phase rather
than a dynamical phase. For the class of low-scale domain-wall scenarios considered here, 
${\cal M}$ naturally lies in the range $10^{-3}\,{\rm eV}$--${\rm eV}$. 
Consequently CMB photons lie at or below the EFT cutoff.
Hence in the phenomenologically relevant regime the propagation 
of CMB photons across the wall is faithfully described by the low-energy theory \eqref{cantra}.

Throughout this work we assume the absence of other ultralight axions or light
fields coupled to electromagnetism. We likewise assume the absence of light electrically charged matter, 
so that electromagnetism is described by free Maxwell theory in each vacuum region. 
Consequently, standard cosmological bounds
associated with light degrees of freedom can be evaded once we consider the
limit in which the axion $\phi$ in Eq. \eqref{cantra} decouples---this being the
only light degree of freedom in the theory besides the photon. In this axion
decoupling limit, the optical activity analyzed below originates entirely from
the vacuum structure encoded in Eq. \eqref{cantra}, rather than from propagating
dark particles. 

The vacuum structure of the theory is determined by unbroken discrete symmetries.
The axion enjoys a discrete shift symmetry,
\be
\frac{\phi}{f_\phi}\rightarrow\frac{\phi}{f_\phi}+n \, ,
\label{thetaax}
\ee
while the quantized dual magnetic field transforms as
\be
{\cal H}\rightarrow{\cal H}+{\cal N}\,{\cal Q}.
\label{thetaH}
\ee
A combined shift leaves the action invariant up to a shift of vacuum
energy. Removing this vacuum energy shift restricts the discrete symmetry
to a single free parameter. Writing ${\cal Q}={\tt q}{\cal M}^2$ with rational ${\tt q} = {\tt r}/{\tt p}$, the vacuum
manifold is the quotient of the integer lattice
$\{n_{\tt vac},{\cal N}_{\tt vac}\}\in\mathbb{Z}^2$ by the sublattice generated by
$({\tt r},{\tt p})$, 
\be
\{n_{\tt vac}+n,{\cal N}_{\tt vac}+{\cal N}\}
\equiv
\{n_{\tt vac},{\cal N}_{\tt vac}\} \, ,
\label{vacuumident}
\ee
with ${\tt r}{\cal N}+{\tt p}n = 0$.

The key physical consequence is that transitions between vacua are discrete and
symmetry protected. A change of vacuum corresponds to varying either $\phi$
or ${\cal H}$ in a way that cannot be undone by local operations. It is these
vacuum transitions---rather than smooth background evolution or local field
values---that control the interaction of photons with the dark sector, and hence
the polarization rotation effects analyzed in the following sections.
	 
\section{Adiabatic Birefringence from Light Axions}

In what follows we treat the CP-odd pseudoscalars appearing in
Eq. \eqref{cantra} as fixed background fields, and the electromagnetic field
$F_{\mu\nu}$ as a probe. This amounts to assuming that the pseudoscalars are
determined by sources entirely decoupled from the electromagnetic sector,
except for their ChernÐSimons coupling to electromagnetism. In this limit the
electromagnetic field equations simplify, and varying the action with respect
to $A_\mu$ yields
\be
\partial_\mu \Bigl\{ F^{\mu\nu} + \frac{\zeta}{6 {\cal M}^2}
\epsilon^{\mu\nu\lambda\sigma}
\bigl({\cal M}^2\frac{\phi}{f_\phi} + {\cal H}\bigr)
F_{\lambda\sigma} \Bigl\} = 0 \, ,
\label{mmon}
\ee
where, as before, we have set $J^\mu = 0$.

Equation~\eqref{mmon} immediately shows that in regions where the combination
$\bigl({\cal M}^2\frac{\phi}{f_\phi} + {\cal H}\bigr)$ is constant, and in the
absence of magnetic monopoles, electromagnetism reduces locally to ordinary
Maxwell theory. The additional term is proportional to
$\epsilon^{\mu\nu\lambda\sigma}\partial_\mu F_{\lambda\sigma}$, which vanishes
identically when the gauge potential can be globally defined,
$\partial_{[\mu}F_{\lambda\sigma]}=0$.

When $\bigl({\cal M}^2\frac{\phi}{f_\phi} + {\cal H}\bigr)$ varies along a
characteristic that propagates the electromagnetic phaseÑi.e.\ along an
ingoing null ray in the geometric-optics limitÑEq.~\eqref{mmon} can induce a
change in the relative orientation of the electric and magnetic fields.
Introducing the auxiliary field strength
\be
\tilde F^{\mu\nu}
=
F^{\mu\nu}
+
\frac{\zeta}{6 {\cal M}^2}
\bigl({\cal M}^2\frac{\phi}{f_\phi}+{\cal H}\bigr)
\epsilon^{\mu\nu\lambda\sigma}F_{\lambda\sigma} \, ,
\ee
it is intuitively clear that the effective linear combination of $\vec E$ and
$\vec B$ rotates as the electromagnetic wave propagates through regions where
the background varies. The central question is which part of this change
corresponds to a genuine physical rotation, and which part may be removed by a
gauge transformation or cancelled by destructive interference. 

To address this, we begin by considering the limit in which the flux ${\cal H}$ is frozen. This
can be achieved by choosing the membrane tension ${\cal T}$ sufficiently large
that the membrane nucleation rate is negligible, so that ${\cal H}$ does not
change over the age of the universe. In this regime we may use the simplified
effective action
\be
S \ni \int d^4 x \Bigl\{
-\frac{1}{4} F_{\mu\nu}^2
- A_\mu J^\mu
- \frac{\zeta}{4! f_\phi}\,
\varphi\,\epsilon^{\mu\nu\lambda\sigma} F_{\mu\nu}F_{\lambda\sigma}
+ \frac12 (\partial\varphi)^2
- V_{\tt I}\!\left(\frac{\varphi}{f_\phi}+\frac{\tilde\theta}{2\pi}\right)
- \frac{{\cal M}^4}{2 f_\phi^2}\,\varphi^2
\Bigr\} \, ,
\label{cantras}
\ee
where we have defined
\be
\varphi = \phi + f_\phi\,\frac{{\cal H}}{{\cal M}^2} \, ,
\qquad
\tilde\theta = \hat\theta - 2\pi\,\frac{{\cal H}}{{\cal M}^2} \, .
\label{varphi}
\ee
This field redefinition isolates the propagating pseudoscalar degree of
freedom while absorbing the effect of the frozen $4$-form flux into a shift
of the effective vacuum angle. The redefined vacuum parameter $\tilde\theta$
encodes the possibility that CP-violating phases arising from ultraviolet
physics are misaligned with those of the dark sector, a natural circumstance
when the axion $\phi$ is of low quality.

\subsection{Axion Walls}

We briefly revisit the analysis of Huang, Harari, and Sikivie
\cite{Huang:1985tt,Harari:1992ea}, but from a perspective that is more suited
to our purposes. The axion background is determined by solving the field
equation obtained from Eq.~\eqref{cantras},
\be
\partial^2 \varphi = - \partial_{\varphi} V_{\tt eff}(\varphi) \, , \qquad
V_{\tt eff}(\varphi)
=
V_{\tt I}\!\left(\frac{\varphi}{f_\phi} + \frac{\tilde\theta}{2\pi}\right)
+ \frac{{\cal M}^2}{2 f_\phi^2}\,\varphi^2 \, .
\label{effaxpot}
\ee
This formulation coincides with the problem studied in
\cite{Huang:1985tt,Harari:1992ea} for photon propagation in the background of a
domain wall generated by a light axion field. The vacua of the theory are
solutions of
\be
\partial_\varphi V_{\tt I}\!\left(\frac{\varphi}{f_\phi} + \frac{\tilde\theta}{2\pi}\right)
+ \frac{{\cal M}^2}{f_\phi^2}\,\varphi = 0 \, .
\label{effaxpotex}
\ee
The periodicity of $V_{\tt I}$ under the discrete shift symmetry
\eqref{thetaax} implies that, when ${\cal H}$ is frozen, neighboring vacua in
the axion lattice are separated by $\Delta\varphi\simeq f_\phi$.

To make the discussion explicit, we adopt the standard harmonic approximation
for the instanton-generated axion potential,
\be
V_{\tt I}
=
\mu^4\!
\left[1-\cos\!\left(2\pi\frac{\varphi}{f_\phi} + \tilde\theta\right)\right],
\label{sineG}
\ee
where $\mu^4$ is the topological susceptibility of the sector generating
$V_{\tt I}$. In this case Eq.~\eqref{effaxpotex} becomes
\be
4\pi^2\mu^4\,\sin(\chi+\tilde\theta)
=
-{\cal M}^4\,\chi \, , \qquad
\chi \equiv 2\pi\frac{\varphi}{f_\phi} \, .
\label{effaxpotexa}
\ee
Solutions of this equation correspond either to true vacua or to maxima of the
effective potential separating them. In the limit of negligible dark-sector
effects, ${\cal M}\to0$, the vacua are located at
$\varphi/f_\phi = n-\tilde\theta/2\pi$ and are separated by
$\Delta\varphi=f_\phi$.

For $0<{\cal M}^4\ll4\pi^2\mu^4$, the solutions can be found perturbatively.
Defining $\chi_n=2\pi n-\tilde\theta$ and expanding about
$\chi=\chi_n+\delta_n$, one finds to leading order
\be
\delta_n
\simeq
-\frac{{\cal M}^4}{{\cal M}^4+4\pi^2\mu^4}
\bigl(2\pi n-\tilde\theta\bigr) \, .
\label{deltas}
\ee
Since $\delta_n\in[0,2\pi]$, multiple distinct vacua exist only if
${\cal M}^4\ll4\pi^2\mu^4$. In the opposite regime the theory admits a unique
vacuum when ${\cal H}$ is frozen\footnote{The minimum then lies generically
far from the endpoints of $[0,2\pi]$, leading to a strong CP problem.}.

The vacuum locations are therefore
\be
\varphi_n
\simeq
\frac{4\pi^2\mu^4}{4\pi^2\mu^4+{\cal M}^4}
\left(n-\frac{\tilde\theta}{2\pi}\right) f_\phi \, ,
\label{shiftmini}
\ee
showing that dark-sector effects both reduce the separation between neighboring
minima,
\be
\Delta\varphi
=
f_\phi
\;\longrightarrow\;
\frac{4\pi^2\mu^4}{4\pi^2\mu^4+{\cal M}^4}\,f_\phi \, ,
\label{shiftphi}
\ee
and lift their degeneracy. The corresponding vacuum energies are non-degenerate, 
\be
V_{\tt eff}(\varphi_n)
\simeq
\frac{{\cal M}^8}{8\pi^2\mu^4}
\left(\frac{\varphi_n}{f_\phi}\right)^2 \, .
\label{minpot}
\ee

Axion domain walls are solitonic field configurations interpolating between
vacua at spatial infinity. Since the energy difference between the true vacuum
and the nearest false vacuum is smallest, we focus on walls connecting those
states. 
Detailed analyses of the interpolating solution appear in
\cite{Huang:1985tt,Vilenkin:2000jqa}. For our purposes it suffices to adopt the
sine-Gordon kink solution \cite{Vilenkin:2000jqa} valid in the regime
${\cal M}^4\ll4\pi^2\mu^4$,
\be
2\pi\frac{\varphi}{f_\phi}+\tilde\theta
=
4\arctan\!\left(e^{m_\phi z/\sqrt{2}}\right) ,
\label{kink}
\ee
where $m_\phi=2\pi\mu^2/f_\phi$ is the axion mass. This solution interpolates
between adjacent vacua with $\Delta\varphi=f_\phi$. The domain wall thickness is
$L\simeq1/m_\phi$, and the maximal gradient at the wall center is
$\partial_z\varphi\simeq m_\phi f_\phi/(\sqrt{2}\pi)$.

We now consider the axion decoupling limit. As in any effective field theory,
the axion decouples when its mass approaches the ultraviolet cutoff,
$m_\phi\to f_\phi$, with $f_\phi$ held fixed. In a Peccei--Quinn UV completion
\cite{Peccei:1977hh,Weinberg:1977ma,Wilczek:1977pj}, $f_\phi$ is the vacuum
expectation value of the radial mode whose phase becomes the axion, and sets
the cutoff of the axion sector. 
In this limit the truncated expansion of $V_{\tt I}$ becomes unreliable and
the wall thickness shrinks to $L\sim1/f_\phi$, while the wall tension,
${\cal T}\simeq\frac23(f_\phi/2\pi)^2 m_\phi$
\cite{Vilenkin:2000jqa}, approaches the cutoff scale,
${\cal T}\sim f_\phi^3$. A consistent effective description therefore requires
\be
\partial_z\varphi < f_\phi^2 \, ,
\label{walldec}
\ee
since variations of order $\Delta\varphi\sim f_\phi$ cannot be resolved over
distances shorter than the cutoff.

\subsection{A Review of Birefringence by Light Axion Domain Walls}

Let us now turn to the propagation of an electromagnetic field in the axion
domain wall background $\varphi(z)$ described above. We review this case in some
detail because it provides a controlled adiabatic limit, which serves as a
baseline for the non-adiabatic and axion-decoupled regimes discussed later.
For a given axion profile $\varphi$, photon dynamics
reduces to standard axion electrodynamics
\cite{Wilczek:1987mv}, with field equations
\be
\partial_\mu \Bigl\{ F^{\mu\nu} + \frac{\zeta}{6 f_\phi}
\epsilon^{\mu\nu\lambda\sigma}  \varphi F_{\lambda\sigma} \Bigl\} = 0 \, ,
\label{mmonax}
\ee
which follow from varying Eq.~\eqref{cantras} and setting $J^\mu = 0$. Since we
also ignore magnetic monopoles, $F_{\mu\nu} = \partial_\mu A_\nu -
\partial_\nu A_\mu$, and hence
$\frac12 \epsilon^{\mu\nu\lambda\sigma} \partial_\mu F_{\lambda\sigma}
= \partial_\mu \, ^* F^{\mu\nu} = 0$, where
$^* F^{\mu\nu} = \frac12 \epsilon^{\mu\nu\lambda\sigma}
\partial_\mu F_{\lambda\sigma}$. Using this identity and substituting for
$\varphi$, the Maxwell equations \eqref{mmonax} in Lorentz gauge reduce to
\be
\partial^2 A^\mu =
g \epsilon^{\mu\nu\lambda\sigma}
\bigl( \partial_\nu \varphi \bigr) \partial_\lambda A_\sigma \, ,
\label{mmonaxpot}
\ee
where we have introduced the shorthand 
\be
g = \frac{\zeta}{3 f_\phi} \, .
\label{geq}
\ee

For propagating electromagnetic waves, residual gauge freedom in Lorentz gauge
may be used to set $A^\mu = (0,\vec A_{\perp},0)$, where $\vec A_{\perp}$ is the
transverse vector field satisfying
$\vec\nabla \cdot \vec A_{\perp} = 0$. Substituting this into
Eq.~\eqref{mmonaxpot} and Fourier transforming in time,
$\vec A_{\perp} \propto e^{-i\omega t}$, after straightforward algebra one
arrives at the matrix equation governing the transverse components:
\be
\Bigl(\bigl(\frac{d}{dz}\bigr)^2 + \omega^2 \Bigr)
\begin{pmatrix}
A^x_\perp \\
A^y_\perp
\end{pmatrix} =
g \omega \partial_z \varphi
\begin{pmatrix}
0 & -i \\
 i & 0
\end{pmatrix}
\begin{pmatrix}
A^x_\perp \\
A^y_\perp
\end{pmatrix} \, .
\label{mateq}
\ee
This equation holds in the rest frame of a very large domain wall, which we
approximate as planar and place at $z=0$. In this frame the translational
homogeneity of the wall implies that $\varphi$ depends only on the coordinate
transverse to the wall, so that $\varphi(\vec x)=\varphi(z)$
\cite{Huang:1985tt,Harari:1992ea}. Also, since we can boost in the $z-t$ plane 
to pick a frame where the 
incident wave is along the normal, $\vec k \parallel \hat z$, 
we have $\vec k \cdot \vec x = k z$, and the
wave phase only changes in the $z$-direction \cite{Huang:1985tt,Harari:1992ea}. 

In the adiabatic regime, $\omega \gg g\,\partial_z\varphi$, the wall is
non-absorptive and incident photons transmit with unit probability to leading
order. Choosing coordinates so that the false vacuum lies at
$z\rightarrow+\infty$ and the true vacuum at $z\rightarrow-\infty$, we may take
all photons to propagate -- i.e. their phases to vary -- along ingoing
characteristics defined by $i\partial_z-\omega=0$.\footnote{We adopt the sign
conventions of \cite{Kaloper:2026slg,Kaloper:2026gib}, in which the initial
``in'' state is located outside the spherical wall at $z\rightarrow+\infty$
and the final ``out'' state at $z\rightarrow-\infty$ in the planar limit. With
this choice the arrow of time corresponds to decreasing $z$, allowing all
integrals to be written as $\int_{-\infty}^{\infty}dz$.} 

Working in the eikonal approximation and projecting onto ingoing modes only,
\[
\Bigl(\bigl(\tfrac{d}{dz}\bigr)^2+\omega^2\Bigr)
=
(\omega+i\partial_z)(\omega-i\partial_z)
\simeq
2\omega(\omega-i\partial_z),
\]
Eq.~\eqref{mateq} reduces to a two-component Schr\"odinger equation with
``imaginary time'' $iz$,
\be
i \partial_z
\begin{pmatrix}
A^x_\perp \\
A^y_\perp
\end{pmatrix} =
H
\begin{pmatrix}
A^x_\perp \\
A^y_\perp
\end{pmatrix} \, , \qquad
H = \omega \unity - \frac{1}{2} g \partial_z \varphi \sigma_2 \, ,
\label{nonmateq}
\ee
where $\sigma_2$ is a Pauli matrix. Mapping wave propagation to a first-order
equation in this way is standard in analyses of axion--photon propagation
\cite{Raffelt:1987im}.

Although the steps above may appear classical, the analysis is properly
interpreted as tree-level quantum mechanics. The transverse fields
$\vec A_\perp$ represent expectation values of the photon vector potential in a
coherent state describing many CMB photons crossing the wall. The eikonal
reduction from the relativistic wave equation \eqref{mateq} to
Eq. \eqref{nonmateq} expresses the absence of mixing between positive- and
negative-frequency modes, justifying the use of quantum-mechanical language.

We emphasize that Eq.~\eqref{nonmateq} is not an exact reduction of the
second order wave equation, but a controlled eikonal approximation. It is
obtained by projecting onto ingoing characteristics and neglecting
backward-propagating modes, and is valid when the interaction terms in $H$ are
ultralocal or parametrically small compared to the photon frequency. When
exploring the behavior of the theory at very high frequencies, we will
therefore need to revert to the second order equation \eqref{mateq}.

Returning to Eq.~\eqref{nonmateq}, the solution is straightforward because the
Hamiltonians at different ``times'' $z$ commute. The evolution operator is
\be
U(z,z_0)
=
{\tt T}\exp\!\left(-i\int_{z_0}^{z}dz\,H(z)\right)
=
\exp\!\left(-i\int_{z_0}^{z}dz\,H(z)\right) .
\label{evol}
\ee
When $\omega \gg |g\,\partial_z\varphi/2|$, the off-diagonal term in $H$ may be
treated perturbatively. Writing
$\Delta\varphi=\int_{-\infty}^{\infty}dz\,\partial_z\varphi$, one finds\footnote{Note that 
$\int_{+\infty}^{-\infty} d(-z) \partial_{-z} \varphi = 
-\int^{+\infty}_{-\infty} dz \partial_{z} \varphi  = - \Delta \varphi$.}
\be
\int_{z_0}^{z}dz\,H =
\omega \Delta z \unity
+ \frac{1}{2} g \Delta\varphi \sigma_2 \, .
\label{intH}
\ee
Substituting this into Eq.~\eqref{evol} yields
\be
U(z,z_0)
=
e^{-i\omega \Delta z}
\begin{pmatrix}
\cos(\frac{1}{2} g \Delta \varphi) & -\sin(\frac{1}{2} g \Delta \varphi) \\
\sin(\frac{1}{2} g \Delta \varphi) & \cos(\frac{1}{2} g \Delta \varphi)
\end{pmatrix} .
\label{evolres}
\ee
The propagated transverse field is therefore
\be
\begin{pmatrix}
A^x_\perp(z) \\
A^y_\perp(z)
\end{pmatrix}
=
e^{-i\omega \Delta z}
\begin{pmatrix}
\cos(\frac{1}{2} g \Delta \varphi) & -\sin(\frac{1}{2} g \Delta \varphi) \\
\sin(\frac{1}{2} g \Delta \varphi) & \cos(\frac{1}{2} g \Delta \varphi)
\end{pmatrix}
\begin{pmatrix}
A^x_\perp(z_0) \\
A^y_\perp(z_0)
\end{pmatrix} .
\label{rotsoln}
\ee

Formal validity of this solution requires $\omega \gg |g\,\partial_z\varphi/2|$.
Using $\Delta\varphi\simeq f_\phi$ and $g=\zeta/(3f_\phi)$ gives
\be
\frac12 g \partial_z \varphi \simeq \frac{\zeta}{6L} \, ,
\ee
where $L\sim1/m_\phi$ is the physical wall thickness, and
$L_{\tt O}=\frac{6}{\zeta}L$ its optical thickness. For wavelengths
$\lambda\ll L_{\tt O}$, the polarization vector undergoes adiabatic rotation by
an angle
\be
\vartheta
=
\frac12 g \Delta\varphi
=
\frac{\zeta}{6} \, ,
\ee
relative to its initial orientation.

In the opposite limit, $\lambda\gg L$, the wall is thin compared to the photon
wavelength. The eikonal equation \eqref{nonmateq} then degenerates as
$\omega\rightarrow0$, and one must revert to the second order equation
\eqref{mateq}. Setting $\omega=0$ yields
$\frac{d^2}{dz^2}\vec A_\perp=0$ for bounded $\partial_z\varphi$, implying
continuity of $\vec A_\perp$ across the wall. Long-wavelength modes therefore
traverse thin walls without polarization rotation.

At this stage the analysis remains entirely within the adiabatic regime and
assumes a smoothly varying axion background. An interesting feature of 
both limits is that the nontrivial dynamics is largely independent of wavelength. 
We will return to this point below. Although this behavior may resemble geometric 
phases in quantum mechanics \cite{Simon:1983mh,Berry:1984jv}, the polarization
rotation here should not be identified with a Berry phase. The evolution
operator \eqref{evol} is path-independent and depends only on endpoint data. If the 
path is closed, the resulting holonomy is trivial. Consequently, BerryÕs
connection and BerryÕs phase vanish identically in these configurations. 
This conclusion hinges crucially on adiabaticity and the absence of
discontinuities in field space. Once these assumptions are relaxed, a different
class of geometric phases becomes relevant.

\section{Chern-Simons and Pancharatnam Phase}

Physically, if the axion is decoupled and its variation is precluded, forcing
$\partial\phi \rightarrow 0$, it is frozen in one of the vacua of
Eqs. \eqref{effaxpotex} or \eqref{effaxpotexa} \cite{Kaloper:2023kua}. In this
regime the dimension-$5$ ChernÐSimons operator in Eq. \eqref{cantra} reduces to a
total derivative. It therefore cannot affect local electromagnetic dynamics,
and photons are oblivious to its presence, provided that the magnetic dual
${\cal H}$ of the top form that monodromizes the axion also remains frozen. By
this we mean that the tension of membranes capable of discharging ${\cal H}$ is
sufficiently large that their nucleation rate is suppressed, so that the
lifetime of ${\cal H}$ exceeds the age of the universe
\cite{Kaloper:2025goq,Kaloper:2026slg,Kaloper:2026gib,Kaloper:2025wgn,Kaloper:2025upu}.

A qualitatively different regime appears if ${\cal H}$ is allowed to change.
Lowering the membrane tension ${\cal T}$ in Eq.~\eqref{cantra} below a critical
value permits domain walls to nucleate within the present Hubble time
\cite{Kaloper:2025goq,Kaloper:2026slg,Kaloper:2026gib,Kaloper:2025wgn,Kaloper:2025upu}.
In that case there is a significant probability that we may inhabit the interior
of a spherical domain wall formed dynamically to discharge ${\cal H}$. From the
action~\eqref{cantra} one then finds ${\cal H}={\cal H}_-+{\cal Q}\,\Theta\!\bigl(r-r(t)\bigr)$, 
where $r$ is the radial coordinate describing the instantaneous radius of the expanding
membrane.

To analyze the effect of such a membrane wall on electromagnetic fields, it is
convenient to consider the large radius limit and approximate the wall as
locally planar, replacing the radial solution by a Cartesian one,
\be
{\cal H} = {\cal H}_- +
{\cal Q}\,\Theta\!\Bigl(n_\mu\bigl(x^\mu-x^\mu_0(t)\bigr)\Bigr) \, ,
\label{step}
\ee
where $x^\mu_0(t)$ parametrizes the wall location, $n_\mu$ is the outward normal,
and $\Theta$ denotes the step function. Here ${\cal H}_-$ is the dual magnetic
flux in the interior vacuum, while
${\cal H}_+={\cal H}_-+{\cal Q}$ characterizes the exterior vacuum. Choosing
$n_\mu$ to point in the $z$-direction gives
\be
{\cal H} = {\cal H}_- +
{\cal Q}\,\Theta\bigl(z-z(t)\bigr) \, .
\label{zH}
\ee

We can continue to approximate the wall with a plane, with the main 
difference compared to the thick light axion walls from the previous
section arising after decoupling the axion, which replaces the term mixing 
polarizations in Eq. \eqref{mateq} by, using Eqs. \eqref{geq} and  \eqref{zH} 
and recalling that ${\cal Q} = {\tt q} {\cal M}^2$,
\be
g \partial_z \varphi \rightarrow g \frac{f_\phi}{{\cal M}^2} \partial_z {\cal H} 
=  g f_\phi \frac{\cal Q}{{\cal M}^2} \delta(z) = 
\frac{\zeta {\tt q}}{3} \delta(z) \, .
\label{params}
\ee
In this regime the
vacuum transition is encoded entirely in the step of ${\cal H}$ across the
interface. At this point the problem is no longer one of photon propagation in a 
slowly varying background, but of scattering across a localized interface separating two distinct vacua.  
This replacement should be viewed not as a limit of a smooth background profile, 
but as the insertion of a localized interaction operator at the interface.

One might be tempted to automatically carry over the results from the thick wall case and replace
\eqref{params} into the Eqs. \eqref{intH} and \eqref{rotsoln}. However 
those results were obtained in the eikonal approximation
valid when positive and negative frequencies are not mixed, and in the 
adiabatic regime where $\omega \gg |g \partial_z \varphi/2|$.
Neither of these conditions is satisfied for thin walls: the $\delta$-function 
does mix positive and negative frequencies, and clearly
invalidates the adiabatic condition.  
Here ``thinÓ does not refer merely to a geometric thickness, but to the 
fact that the interaction is localized on scales that invalidate adiabatic transport altogether. 
Accordingly, the appropriate description of photon dynamics is no longer a 
first-order transport equation, but a genuine 
scattering problem governed by the second order wave equation.  

For these reasons the problem of waves traversing the wall was analyzed in
Refs. \cite{Kaloper:2026slg,Kaloper:2026gib} by treating the wall as the 
boundary conditions for electromagnetic fields
on the interface, similar to Fresnel refraction in optics. Here, we 
will use a more general approach which will provide
deeper insight into the microscopic dynamics of wall crossing, 
reproducing precisely the results of  \cite{Kaloper:2026slg,Kaloper:2026gib} 
in the limit where the mixing strength in \eqref{params} is small.  

To treat this interface scattering exactly, we rewrite the second order wave 
equation as a Lippmann--Schwinger integral equation, 
which makes reflection, transmission, and unitarity manifest.  
In this setting the Lippmann--Schwinger formulation is not an approximation 
scheme, but the natural language for interface scattering problems.
To this end, we retreat to the second order equation \eqref{mateq} 
and solve it directly. Substituting \eqref{params} into 
\eqref{mateq} yields the thin wall second order propagation equation, which is exact: 
\be
\Bigl(\bigl(\frac{d}{dz}\bigr)^2 + \omega^2 \Bigr)\vec A_\perp
=\frac{\zeta{\tt q}}{3}\,\omega\,
\delta(z)\,
\sigma_2\,\vec A_\perp\,.
\label{mateqsm}
\ee
To solve it we invert the differential operator using the   
outgoing Green's function $-\frac{i}{2\omega}e^{i\omega|z|}$ for \eqref{mateqsm}. 
This replaces the 
Schr\"odinger differential equation \eqref{mateqsm} with the 
completely equivalent Lippmann-Schwinger integral equation 
\be
\vec A_\perp(z) =
e^{-i\omega z}\,\vec A_\perp(+\infty) - i\,\frac{\zeta{\tt q}}{6}\,\sigma_2
\int_{-\infty}^{\infty}dz'\,
\delta(z') \,
\vec A_\perp(z')\,e^{i\omega |z-z'|}\, , 
\label{borneq}
\ee
that propagates the two--component wavefunction $\vec A_\perp(z)$ from 
$+\infty$ to $-\infty$, realizing our boundary condition
where a wave comes in ``outside" of the wall and moves across 
it toward an observer ``inside". The form of the integrand in
\eqref{borneq} shows explicitly that the $\delta$-term mixes 
positive and negative frequencies, as it produces a reflected
wave traveling opposite to the incident one, back to $+\infty$.  
This frequency mixing is not a small correction but a structural feature of the interface interaction. 

This equation can be integrated extremely simply, thanks to the $\delta$-function.  
The exact solvability of the scattering problem is a special consequence 
of the $\delta$-localized interaction. For any extended wall profile the solution 
would remain nonlocal. Indeed,
\be
\vec A_\perp(z) =
e^{-i\omega z}\,\vec A_\perp(+\infty) - i\,\frac{\zeta{\tt q}}{6}\,\sigma_2 \,
\vec A_\perp(0)\,e^{i\omega |z|}\, , 
\label{borneqs}
\ee
and so, evaluating the solution at $z=0$, we can solve for 
$\vec A_\perp(0)$ in terms of $\vec A_\perp(+\infty)$. Using shorthand
notation $\hat \sigma = \zeta {\tt q}/6$ and $\vartheta = \tan^{-1}(\hat \sigma)$, 
after straightforward algebra with Pauli matrices,
\be
\vec A_\perp(0) = \frac{1}{\sqrt{1+\hat \sigma^2}} 
e^{- i\,\vartheta\,\sigma_2} \,
\vec A_\perp(+\infty) \, , 
\label{a0}
\ee
Substituting into the solution \eqref{borneqs} yields 
the {\it exact solution} of \eqref{mateqsm} describing an incoming
wave refracting on the wall,
\be
\vec A_\perp(z) =
\Bigl(e^{-i\omega z}\, \unity - i\,\sin(\vartheta)\, e^{i\omega |z|} \, 
e^{- i\,\vartheta\,\sigma_2} \, \sigma_2  \Bigr) \vec A_\perp(+\infty) \, , 
\label{borneqsa}
\ee
which for $z>0$ describes a linear combination of the incident 
and reflected wave, and for $z<0$ the transmitted wave. It is
illuminating to write these explicitly, 
\be
\vec A_\perp(z) =
\begin{cases}
~ e^{-i\omega z}\, \vec A_\perp(+\infty) - i\,\sin(\vartheta)\, e^{i\omega z} \, 
e^{- i\,\vartheta\,\sigma_2} \, \sigma_2 \vec A_\perp(+\infty)  \, ,
\hfill  ~~~~~  z>0  \, ;\\
\\
 ~ e^{-i\omega z} \, \cos(\vartheta)\, 
 e^{- i\,\vartheta\,\sigma_2} \, \vec A_\perp(+\infty)  \, , 
 \, ~~~~~~~~~~~~~~~~~~~~~~~~~~~~~ \, z< 0  \, . 
\end{cases}
\label{wavessum}
\ee
Equation \eqref{wavessum} thus provides the complete and exact description of 
photon scattering across the wall, including reflection, transmission, and polarization rotation.  

These solutions elegantly reveal the underlying physics. 
First of all we see that indeed the interactions do produce frequency 
mixing and reflected wave in general. However, when the 
mixing is weak, $\hat \sigma \ll 1$ and so $\vartheta \ll 1$, the reflected wave 
amplitude vanishes as $\sin(\vartheta) \rightarrow 0$,
restoring the range of validity of the eikonal approximation we used in the 
previous section. Secondly, the unitarity of the solution is
maintained since the transmitted, reflected and incident amplitudes add up, 
$\sin^2(\vartheta) + \cos^2(\vartheta) = 1$, as is clear
from \eqref{wavessum}. This exactly confirms the perturbative arguments for 
unitarity deployed in \cite{Kaloper:2026slg,Kaloper:2026gib}.
Third, both the polarization of transmitted and reflected waves are rotated
by the same angle $\vartheta$; this is indeed a precondition for the 
wave intensity conservation at the wall, as required by unitarity.  

Finally, the polarization orientation changes discretely at the thin wall 
for all wavelengths of photons below the cutoff of the theory
describing the wall at low energies. This polarization rotation precisely 
meets the Pancharatnam phase definition 
\cite{Pancharatnam:1956url,Pancharatnam:1956url2}.  
Because the interaction is localized at the interface, the polarization 
change is finite and discontinuous, making it natural to interpret the resulting phase 
as a Pancharatnam phase associated with state overlap across the wall. 
To expose the geometric content of this polarization rotation, 
it is natural to work in the circular polarization basis, 
where the wall interaction diagonalizes.  
In this problem the non-trivial Pancharatnam phase is exposed only in helicity eigenstates. 
We rewrite our solution \eqref{wavessum} in the 
circularly polarized basis 
\be A_\pm(z)
= \frac{1}{\sqrt{2}}\Bigl(A_\perp^x(z) \pm i A_\perp^y(z)\Bigr) \, .
\label{circcomp}
\ee
In terms of these components $\sigma_2$ is diagonal, $\sigma_2 A_\pm = \pm A_\pm$. Now,
consider an incoming wave of a definite circular polarization,
\be
A_\pm(z>0) = e^{-i\omega z}\,A_\pm(+\infty) \, ,
\qquad \qquad
A_\mp(+\infty)=0 \, .
\ee
The transmitted part of \eqref{wavessum} for $z<0$ is
\be
A_\pm(z<0) = \cos(\vartheta)\, e^{\mp i\vartheta}\,
e^{-i\omega z}\, A_\pm(+\infty),
\qquad
A_\mp(z<0)=0.
\label{transcirc}
\ee
Hence the two circular polarization components cross the wall independently from each other, 
acquiring a helicityÐdependent phase factor $e^{\mp i\vartheta}$. As we explained above, the
normalization $\cos(\vartheta)$ accounts for the reduction of transmitted
intensity due to partial reflection and does not affect relative phases.  

The Pancharatnam phase is a ray-space quantity: reflection affects amplitudes 
but not the geometric phase extracted from normalized state overlap. 
The polarization rotation encoded in the transmitted wave can now be interpreted 
as a geometric phase, extracted directly from the overlap 
of incoming and outgoing polarization states. 
To extract this Pancharatnam phase, we compare the incoming and outgoing
polarization states after normalizing away this real amplitude factor. The
incoming polarization wavefunction is specified by the complex amplitude
$e^{-i\omega z} A_\pm(+\infty)$, and the transmitted polarization is specified by 
$\tilde A_\pm(z<0)
= \cos(\vartheta)\,e^{\mp i\vartheta}\,A_\pm(+\infty)$. 
The Pancharatnam phase is defined as the phase of the overlap between the
incoming and outgoing polarization states, divided by its magnitude 
to properly normalize the overlap to unity. 
In the present case this
reduces to the relative phase between the complex amplitudes,
\be
e^{i\gamma_{\rm P}^{(\pm)}}
=
\frac{A_\pm(+\infty)^\ast\,\tilde A_\pm(z<0)}
{|A_\pm(+\infty)^\ast\,\tilde A_\pm(z<0)|}
=
e^{\mp i\vartheta} \, .
\label{panchph}
\ee
Hence the Pancharatnam phase acquired by the two helicity components is
\be 
\gamma_{\rm P}^{(\pm)} = \mp\,\vartheta \, . 
\label{panfinal}
\ee
Our derivation makes explicit that the Pancharatnam phase arises directly from
the helicityÐdependent phase acquired by the circular components of the
transverse electromagnetic field upon crossing the ChernÐSimons wall, irrespective of
the adiabatic approximation, other auxiliary bases or the interaction strengths, that can cause 
reflection, as long as everything occurs in the effective theory below the cutoff.  
No Berry connection, curvature, or adiabatic parameter space transport enters at any stage. 
Unlike a Berry phase, a Pancharatnam phase does not require transport
around a closed loop in parameter space. It is defined directly from the overlap of 
the initial and final polarization rays. 
The geometric phase is read off directly from the exact scattering
solution \eqref{wavessum}.

The interactions do affect the magnitude of the polarization rotation angle. In the limit of weak
coupling of the photon to the wall, when $\hat \sigma \ll 1$, the rotation angle is 
\be
\Delta\vartheta = \frac{\zeta{\tt q}}{6} \, ,
\label{panchrot}
\ee
which precisely reproduces the adiabatic limit induced by a very light axion variation of the previous
section when ${\tt q}=1$. 
The sign of the rotation is determined by the orientation of the vacuum transition. 
Reversing the direction of wall crossing changes $\Delta\Theta \rightarrow -\Delta\Theta$ 
and therefore $\Delta\vartheta \rightarrow -\Delta\vartheta$. Thus interfaces crossed in 
opposite directions contribute with opposite signs to the total rotation. 
However, as we see, the full physical picture is far more intricate. Even in the
adiabatic limit, the actual rotation angle is not really there because of the light axion by itself, but because of the
complex vacuum structure that enable that axion to interpolate between the vacua, 
and because the photon transits from one vacuum region to another. Further the exact same result can be
reproduced by the thin walls without any light axions, as already noted in \cite{Kaloper:2026slg,Kaloper:2026gib}.

We recall that in the adiabatic limit, the polarization rotation 
is frequency independent. This persists in the thin wall limit as well.  
Within the regime where the effective coupling $\zeta$ may be treated 
as approximately frequency independent, 
perhaps the easiest way to see why this curious feature emerges is 
directly from the second order differential equation \eqref{mateqsm}. 
Dividing this equation by $\omega^2$ and changing 
variable to $\xi = \omega z$ yields
\be
\Bigl(\bigl(\frac{d}{d\xi}\bigr)^2 + 1 \Bigr) 
\vec A_\perp = 
\frac{\zeta  {\tt q}}{3} \delta(\xi)
\sigma_2 \vec A_\perp \, ,
\label{mateqtop}
\ee
with frequency completely absorbed below the cutoff ${\cal M}$, when we approximate the wall with a 
$\delta$-function, and as long as the coupling $\zeta$ is approximately frequency-independent. This frequency 
independence is the crucial mechanism that allows us to identify the polarization rotation as a
symmetry-protected probe of the topology of EFT vacua, and
follows from $1$-form symmetry in the photon-wall system which we will discuss later.  

Since the photon-wall system is so simplified for thin walls, one might ask if smearing the wall changes the 
picture in salient ways that obstruct the symmetry-protected geometric interpretation of polarization rotation. 
In short, the answer is no: while smearing the wall 
changes some details of photon wall crossing, the interpretation of 
polarization rotation as the Pancharatnam phase remains robust as long
as the frequencies remain well below the cutoff of the complete photon-axion-wall theory in \eqref{cantra}. 

To show this, let us thicken the wall by introducing a finite wall-width scale $m_w$, 
which in general is below the ultraviolet cutoff ${\cal M}$ of the effective theory. 
In axionic realizations one typically expects $m_w \sim m_\phi$, whereas the EFT cutoff may 
be parametrically larger. The discussion below illustrates robustness of the geometric phase 
and should be viewed as a probe of finite wall thickness rather than of ultraviolet completion. 
We can model this very simply by replacing the $\delta$-function in \eqref{mateqsm} by a 
smooth distribution $\delta_{m_w}$ that smears the Dirac $\delta$-function.

In this case the Lippmann-Schwinger equation \eqref{borneqs} changes to
\be
\vec A_\perp(z) =
e^{-i\omega z}\,\vec A_\perp(+\infty) - i\,\frac{\zeta{\tt q}}{6}\,\sigma_2
\int_{-\infty}^{\infty}dz'\,
\delta_{m_w}(z') \,
\vec A_\perp(z')\,e^{i\omega |z-z'|}\, , 
\label{borneqsm}
\ee
and we cannot integrate it in closed form. But that is not necessary; 
using circular polarizations, we diagonalize the matrix, and reduce \eqref{borneqsm} to two decoupled
wave equations where we normalize both initial polarizations to unity at $z \rightarrow \infty$,
\be
\psi_\pm(z) =
e^{-i\omega z} \mp 
\,\frac{\zeta{\tt q}}{6}\, 
\int_{-\infty}^{\infty}dz' \,
\delta_{m_w}(z') \,
e^{i\omega |z-z'|} \psi_\pm(z') \, . 
\label{borneqpol}
\ee
Now we can separately consider the transmission and reflection by taking the limits $z \rightarrow \mp \infty$ 
for transmission and for reflection coefficients, respectively. We 
compute them by the first order term in the Born approximation to
\eqref{borneqpol}. With our boundary conditions, after some straightforward algebra, using 
$z \rightarrow \mp \infty$, $|z-z'| = \mp(z' -z)$, this finally yields 
\be
\delta t_\pm = \mp 
\,\frac{\zeta{\tt q}}{6}\, e^{-i \omega z} \, \int_{-\infty}^{\infty}dz'\,
\delta_{m_w}(z')   \, ,  ~~~~~~~~ \delta r_\pm = \mp 
\,\frac{\zeta{\tt q}}{6}\, e^{i \omega z} \, \int_{-\infty}^{\infty}dz'\,
\delta_{m_w}(z')  \, e^{- 2i \omega z'} \, .
\ee
The dependence of the integrands on $\omega$ is the key issue here. In the case of the transmission
coefficient, the integral adds to unity, since $\delta_{m_w}(z')$ 
is normalized to unity, and this occurs for {\it any} value of
frequency $\omega$ in this approximation. In contrast, 
the integrand in the reflection integral depends on frequency explicitly, 
and the limits $\omega \ll m_w$ and $\omega \gg m_w$ lead to different results.
In the former case, since the frequency is low, we 
can ignore the phase factor $e^{- 2i \omega z'}$ to leading order. 
Thus in this limit, the integral is again unity. Contrasting it, when $\omega \gg {m_w}$, 
the phase term oscillates rapidly and the integral converges to zero by Riemann-Lebesgue lemma. 
We stress that this result is completely 
independent of the form of the smearing function $\delta_{m_w}(z)$. 
Thus we find
\ba
\delta t_\pm &=& \mp
\,\frac{\zeta{\tt q}}{6}\, e^{-i \omega z} \, , ~~~~~~ \delta r_\pm = \mp 
\,\frac{\zeta{\tt q}}{6}\, e^{i \omega z} \, ,  ~~~~~~ \omega \ll {m_w} \, , \nonumber \\
\delta t_\pm &=& \mp 
\,\frac{\zeta{\tt q}}{6}\, e^{-i \omega z} \, , ~~~~~~ \delta r_\mp = 0 \, , \, ~~~~~~~~~~~~~~~~ \omega \gg {m_w}  \, .
\label{tandr}
\ea
In the high frequencies limit there is no reflection, but optical activity persists and the transmitted waves 
retain their Pancharatnam phases.  

This changes when as we approach the cutoff we 
recover the frequency dependence of $\zeta$ which we discussed above. 
Operationally, our Schr\"odinger potential came from a 
wall-borne dimension-5 operator, which must include a power of frequency in the
{\it numerator}, as is evident on the right-hand side of Eq. \eqref{mateq}. For small frequencies the power 
of $\omega$ brought in by the Chern-Simons terms cancels the power of $\omega$ in the
denominator of the outgoing Green's function $-i e^{i \omega|z-z'|}/\omega$ 
which we used in computing the transmission amplitude.
As a result the phase shift remains a robust feature for even high 
frequency modes, effectively enabling them to probe the field space topology.

However, Eq. \eqref{cantra} arises as a low energy effective 
theory, by integrating out heavy fields with 
anomalous coupling to electromagnetism. 
The parameter $\zeta$ must be scale dependent, and when frequencies rise 
above the cutoff the effective coupling decreases with $\omega$. 
Our example in Eq. \eqref{zetaeq} entails
$\zeta \simeq {\cal M}^2/\omega^2$, which would modify 
the leading order Born contribution to $t_\pm$ in the second line of Eq. \eqref{tandr} to 
$\delta t_\pm \rightarrow \mp \frac{{\cal M}^2}{\omega^2} \,\frac{\zeta{\tt q}}{6}\, e^{-i \omega z}$, 
suppressing the phase shift for high frequency electromagnetic waves,
and making the wall essentially transparent at energies much higher than ${\cal M}$. 
While the precise form of dependence of $\zeta$ on $\omega$ 
is detail-dependent, the expected general trend 
is the same. This is the situation we envisioned in the earlier papers 
\cite{Kaloper:2025goq,Kaloper:2026slg,Kaloper:2026gib}.

\section{Higher-Form Symmetry, Chern--Simons Descent and Topological Protection}

Identifying the mechanism of polarization rotation with the Pancharatnam phase of electromagnetic wave
propagation in axion electromagnetism invites a natural interpretation in terms of a mixed higher-form
symmetry structure \cite{Gaiotto:2014kfa,Hidaka:2020iaz,Brennan:2020ehu}
between electromagnetic $1$-form symmetry and the discrete vacuum structure of the dark sector.
This interpretation emerges by resorting to the Chern--Simons descent structure of the monodromized axion
coupling to electromagnetism $\propto \Theta F\wedge F$, where from Eq.~\eqref{cantra}
\be
\Theta =   \bigl(\sqrt{\cal X}\frac{\phi}{f_\phi} + {\cal H} \bigr) \, ,
\label{0form}
\ee
is the monodromized axion field which may include both smooth variation (if/when the axion is light) and
discrete jumps (if/when the dual magnetic form ${\cal H}$ discharges). This quantity is constant outside of a defect,
in regions of the fixed $\Theta$-vacuum. Note that we do not need any background
cosmological time variation of the axion to induce variation of $\Theta$, 
but only codimension-$1$ domain walls which source
$\Theta$. The role of the descent structure is to fix the normalization of the wall-supported
Chern--Simons operator, which in turn uniquely determines the unitary matching condition for photon
polarization across the interface. The polarization rotation which we calculate in the EFT is completely fixed by the
normalization of the wall Chern--Simons theory. It is therefore a
symmetry-protected geometric phase associated with crossing a defect carrying quantized
topological charge. The symmetry analysis is performed at the level of the 
infrared CP-odd effective theory, whose topological couplings are inherited 
from anomaly data and are therefore radiatively stable.

At energies well below the dark sector cutoff ${\cal M}$, after we integrated out dark degrees of freedom,
whose masses would naturally be around ${\cal M}$, the dark sector can influence electromagnetism by
the CP-odd interaction
\be 
S_{\rm CS}
=
- \frac{\zeta}{4!{\cal M}^2}
\int d^4x\,
\Theta(x)\,
\epsilon^{\mu\nu\lambda\sigma}F_{\mu\nu}F_{\lambda\sigma}
= 
\frac{\zeta}{6{\cal M}^2}
\int \Theta\, F\wedge F \, , 
\label{CSbulk_rewritten}
\ee
where $F = \frac12 F_{\mu\nu} dx^\mu \wedge dx^\nu$ is the 
electromagnetic field strength $2$-form. Notice that in this
regime, with energies below the dark sector cutoff, 
we are also exponentially below the masses of all real electrically
charged particles. Hence, in the absence of boundary-induced 
effects, conventional electromagnetism would be trivial
without moving charged sources. The coupling \eqref{CSbulk_rewritten} provides precisely such boundary-induced
effects: domain walls support localized interactions that impose nontrivial matching conditions on electromagnetic
fields across the interface. The term \eqref{CSbulk_rewritten} is not a total derivative once $\Theta$ is
space-dependent. Using the standard Chern--Simons descent identity,
$\Theta F\wedge F = d(\Theta \, A\wedge F) - d\Theta \wedge A\wedge F$. 
After dropping the total derivative, the physically relevant contribution is entirely controlled by gradients of
$\Theta$. In regions where $\Theta$ is constant, the interaction is a total derivative and does not affect local
electromagnetic dynamics.

In contrast, on the wall $d\Theta = \Delta \Theta\, \delta_{\rm wall}$, regardless of whether the
$1$-form $\delta_{\rm wall}$ is discrete or smooth. This yields a
codimension-$1$ Chern--Simons interaction,
\be 
S_{\rm wall} = - \frac{\zeta \Delta \Theta}{6{\cal M}^2}
\int_{\rm wall} A\wedge F \, .
\label{wallCS}
\ee
The wall-supported interaction means that the wall is a charged codimension-$1$ defect for
electromagnetic $1$-form symmetry in the low-energy effective theory, as anticipated above.
Its physical role is to impose a symmetry-constrained interface operator whose normalization
is fixed by Chern--Simons descent. 
The region where $d\Theta \ne 0$ is the support of the
descent of the topological density $F\wedge F$ triggered by a discontinuity in $\Theta$.

To make this more specific we define the defect-supported $3$-form current
\be 
J^{(3)} \equiv d\Theta \wedge F \, ,
\label{defectcurrent}
\ee
which is conserved off the wall in our EFT due to $dF = d^2 \Theta =0$ away from the defect.
From the higher-form symmetry perspective, $J^{(3)}$ encodes the conserved defect charge
associated with electromagnetic $1$-form symmetry in the presence of domain walls.
Since the support of this current is entirely localized on the wall, it represents
a purely interface degree of freedom. In terms of it we can rewrite the Chern--Simons
interaction as
\be 
S_{\rm CS} = - \frac{\zeta}{6{\cal M}^2} \int  J^{(3)} \wedge A \, .
\label{Sdesc}
\ee
Although the scalar $\Theta$ is hidden in the wall current $J^{(3)}$, it is its variation that turns $J^{(3)}$ on
since $d\Theta = \Delta \Theta\, \delta_{\rm wall}$ implies $J^{(3)} = \Delta \Theta\, \delta_{\rm wall} \wedge F$.
Thus the coupling of $J^{(3)}$ to $A$ -- the photon incoming from the bulk -- 
completely fixes the normalization of the wall-supported Chern--Simons operator.
Its value is insensitive to details of bulk physics below the cutoff ${\cal M}$, provided
the wall topology is unchanged. Since the walls separate domains with different values of 
the scalar ``flux'' $\Theta$, defined
on the vacuum manifold labeled by discrete data $\{n_{\rm vac},{\cal N}_{\rm vac}\}$, the 
photons crossing the wall pick up the fixed polarization 
twist set by $\Delta\Theta = {\cal Q} = {\tt q}\,{\cal M}^2$.
Indeed, from the descent perspective, the wall carries a quantized defect charge,
\be
Q_{\rm wall} = \int J^{(3)} = \int_{\Sigma_3} d\Theta \wedge F \, ,
\ee
which is invariant under smooth deformations of the wall and defines a superselection
structure in the low-energy EFT. Local operators cannot change this charge. It can
disappear only when the wall itself is eliminated by ultraviolet effects.
This completely fixes the polarization rotation which we calculate in the EFT by tying it to
the wall charge in the wall Chern--Simons theory. 

Indeed, combining the parameters in the equations, 
we see that we can rewrite the coupling \eqref{Sdesc} 
as 
\be 
S_{\rm wall} = - \frac{\zeta {\tt q}}{6} \int_{\rm wall} F \wedge A \, .
\label{walloCS}
\ee
and recognize the coefficient as precisely the Pancharatnam phase 
we calculated above. Since $F \wedge A$ is CP-odd, the
action will pick up a sign depending on which precise photon 
polarization we are describing. If we consider an incoming 
linearly polarized wave, this then implies that the transmitted 
wave will have its polarization rotated by the angle controlled by
the coefficient in \eqref{wallCS}. 

The wall Chern--Simons term should be understood as defining a localized interface
operator rather than as contributing to phase accumulation along the photon
trajectory. Because the interaction is supported on a codimension-$1$ surface
and is bilinear in the electromagnetic field, its variation does not generate a
force or potential, but instead imposes matching conditions that relate the
electromagnetic fields on the two sides of the wall. Crossing the interface
therefore implements a finite canonical transformation of the transverse
electromagnetic degrees of freedom, mapping incoming polarization states to
outgoing ones by a fixed linear operator whose normalization is fixed by the
Chern--Simons descent structure.

This is easily made explicit in the planar wall approximation, which we discussed earlier. 
The wall action \eqref{walloCS} adds a $\delta$-function contribution to the second--order wave
equation for the transverse modes, which we repeat here,
\be
\Bigl(\partial_z^2 + \omega^2\Bigr)\vec A_\perp =
\frac{\zeta{\tt q}}{3}\,\omega\,\delta(z)\,\sigma_2\,\vec A_\perp \, ,
\label{mateqwall}
\ee
which enforces continuity of $\vec A_\perp$ across the interface while inducing a
finite jump in its conjugate momentum, and where $\sigma_2$ merely takes stock of chirality 
in a general basis. Then integrating \eqref{mateqwall} across an
infinitesimal neighborhood of the wall yields a 
matching condition of the form valid for small $\vartheta$, 
\be
\vec A_\perp(0^-) = U_{\rm wall}\,\vec A_\perp(0^+),
\qquad
U_{\rm wall} = \exp\!\bigl(- i\,\vartheta\,\sigma_2\bigr),
\qquad
\vartheta=\frac{\zeta{\tt q}}{6}.
\label{Uwalleq}
\ee
The wall acts as a unitary operator on the $2D$ polarization
Hilbert space. In the helicity basis, where $\sigma_2$ is diagonal, the circular
polarization states are eigenstates of $U_{\rm wall}$,
\be
U_{\rm wall}\,|\pm\rangle = e^{\mp i\vartheta}\,|\pm\rangle ,
\label{helicityjump}
\ee
so that each helicity acquires a definite phase upon crossing the interface,
while a linearly polarized state is rotated by an angle $\vartheta$. This is exactly 
the approach used in \cite{Kaloper:2026slg,Kaloper:2026gib}.

From this perspective, the polarization rotation is not generated by integrating
a connection along the photon worldline, but arises as the eigenphase of a finite
unitary matching operator localized at the interface. The geometric phase is
extracted from the overlap of incoming and outgoing polarization states after
the action of $U_{\rm wall}$, precisely in the sense of Pancharatnam. The role of
the Chern--Simons descent is to fix the normalization of this operator, while the
emergent $1$-form symmetry ensures that its action is protected from smooth 
microscopic deformations of the wall profile. 

It is useful to view this mechanism as the relativistic analogue of a thin
optical element, such as a wave plate \cite{Born:1999ory}. A wave plate does not
induce birefringence through propagation, but instead applies an instantaneous
canonical transformation that mixes polarization components at a single
location. Likewise, the wall-localized Chern--Simons interaction implements a
polarization-dependent phase shifter embedded in spacetime, whose observable
effect depends only on the integrated strength of the interface interaction and
not on its microscopic thickness.

It is instructive to contrast this mechanism with the conventional narrative of
axion-induced birefringence. In axion electrodynamics, polarization rotation
arises from spacetime variation of a light scalar and is commonly interpreted as
the dynamical accumulation of a phase proportional to
$\int \dot{\phi}\,dt$ along the photon trajectory. While the final effect may
appear similar, it is highly sensitive to the detailed evolution of the axion
background and requires axions lighter than $10^{-28}\,\mathrm{eV}$, which are
already constrained by cosmology, see e.g. \cite{Amendola:2005ad,Hlozek:2014lca}. By
contrast, in the present framework the polarization rotation is a
symmetry-protected geometric quantity in the low-energy EFT, fixed by the
quantized normalization of the wall Chern--Simons operator. The effect is
localized at vacuum interfaces, frequency-independent below the cutoff, and does
not accumulate continuously with propagation distance, requiring only the
existence of distinct vacuum domains separated by walls supporting
electromagnetic Chern--Simons interactions
\cite{Kaloper:2025goq,Kaloper:2026slg,Kaloper:2026gib}.

A natural question is how such walls arise in the dark sector and how a coupling
between $\Theta$ and $F \wedge F$ is generated. As discussed earlier, this can
occur through integrating out heavy pseudoscalars in ultraviolet completions of
the wall physics. From the viewpoint of higher forms and symmetries, a simple
realization arises if the dark sector mixes with QCD and electromagnetism through
two axion portals, as in Discretely Evanescent Dark Energy models based on
nonabelian gauge dynamics \cite{Kaloper:2025goq}. At energies well below all
confinement scales, after integrating out all massive degrees of freedom,
neither the dark sector nor QCD appear as propagating gauge theories. Instead,
each sector admits a dual description in terms of $3$-form gauge fields \`a la
L\"uscher \cite{Luscher:1978rn},
\be
{\rm Tr}\Bigl(G_{\tt QCD} \wedge G_{{\tt QCD}} \Bigr) = 
dC^{(3)}_{\tt QCD} = {\cal G}_{\tt QCD} \, , \qquad \qquad 
{\rm Tr}\Bigl(G_{\tt dark} \wedge G_{{\tt dark}} \Bigr) = 
dC^{(3)}_{\tt dark} = {\cal G}_{\tt dark} \, ,
\label{luscherfroms}
\ee
which encode fluctuations of vacuum energy density and $\theta$-like sectors
rather than propagating gluons or dark photons. The electromagnetic sector
remains described by a dynamical $1$-form gauge field $A$ with field strength
$F=dA$, while the axionic sector is captured by pseudoscalar fields $a_i$ with
compact shift symmetries arising from ultraviolet anomalies and membrane
physics.

Once heavy states charged under both nonabelian gauge groups are integrated out,
the infrared theory is no longer a direct product of independent sectors.
Instead, it is constrained by a coupled topological structure that ties together
axion shift symmetries, $3$-form gauge symmetries, and electromagnetic topology.
Schematically, the resulting low-energy theory is described by
\ba
{\cal L}_{\tt IR} &=& -\frac{1}{2}F \wedge ~^{*}F - 
\frac{1}{48} {\cal G}_{{\tt QCD}}\wedge ~^{*} {\cal G}_{{\tt QCD}}  
- \frac{1}{48} {\cal G}_{{\tt dark}}\wedge ~^{*} {\cal G}_{{\tt dark}}  \nonumber \\
&& +\sum_{i=1}^{2} \Bigl[ \frac{1}{2}d a_i \wedge ~^{*} da_i +
a_i \Bigl(\left(k_i {\cal G}_{\tt QCD} + \ell_i {\cal G}_{\tt dark} \right) 
+ \frac{1}{8\pi^2} c_i F \wedge F \Bigr) \Bigr] \, .
\label{IRL}
\ea
The mixed topological couplings prevent factorization of the symmetry structure
and enforce compensating transformations among axions, $3$-form gauge fields,
and electromagnetic topology. Gauge invariance of the electromagnetic sector
under $A \rightarrow A + d\lambda$ would normally leave the remaining sectors
untouched; however, the mixed couplings require correlated transformations in
the axion and $3$-form sectors. While the $3$-forms are invariant under
$C^{(3)}_i \rightarrow C^{(3)}_i + d \Lambda_i^{(2)}$ and the axions under
shift symmetries $a_i \rightarrow a_i + \alpha_i$, these transformations are
twisted once the mixed anomalies are taken into account,
\be
\delta C_{\tt QCD} = - \sum_i k_i \alpha_i \, \Upsilon^{(3)} \, , \qquad
\delta C_{\tt dark} = - \sum_i \ell_i \alpha_i \,\Upsilon^{(3)} \, ,
\label{twisted}
\ee
where $\Upsilon^{(3)}$ is a fixed background $3$-form encoding the electromagnetic
topological sector through its descent relation
\be
d\Upsilon^{(3)} = \frac{1}{8\pi^2} F \wedge F \, .
\label{Xidef}
\ee
This constraint expresses the obstruction to factorizing axion shift symmetries
from electromagnetic topology. As a result, the axion shift symmetry, the QCD and
dark $3$-form symmetries, and emergent electromagnetic topology combine into a
nontrivial higher-group structure, in which gauge transformations at one level
act nontrivially on higher-form symmetry data at another.

\section{Multiple Interfaces and Observable Birefringence}

An important consequence of the interface picture developed above concerns
the observational structure of the birefringence signal. Since in principle photons could
encounter multiple vacuum interfaces between the last scattering surface and
the observer, it is natural to ask whether successive wall crossings
accumulate stochastically, generating a random walk of polarization
angles and a nontrivial angular power spectrum. In the class of theories
considered here, this is not the generic outcome. This feature is a key observational 
input of the mechanism explored here.

To see this, consider a sequence of vacuum transitions along a given line
of sight. Each interface contributes a polarization rotation,
\be
\Delta\vartheta_i
=
\frac{\zeta}{6{\cal M}^2}\,
\Delta\Theta_i \, ,
\qquad
\Delta\Theta_i
=
\Theta_{i+1}-\Theta_i \, .
\label{walljump}
\ee
with $\Theta$ defined in \eqref{0form}. 
For emphasis again, we remind the reader that reversing the direction of wall crossing 
changes $\Delta\Theta_i \rightarrow - \Delta\Theta_i$; thus $\Delta\vartheta_i \rightarrow - \Delta\vartheta_i$.

The total rotation is therefore
\be
\Delta\vartheta_{\rm tot}
=
\frac{\zeta}{6{\cal M}^2}
\sum_i \Delta\Theta_i \, .
\label{manywalls}
\ee
However the sum immediately telescopes,
\be
\Delta\vartheta_{\rm tot}
=
\frac{\zeta}{6{\cal M}^2}
\Bigl(
\Theta_{\rm obs} -
\Theta_{\rm emit}
\Bigr)\, .
\label{endpointrotation}
\ee
Thus all intermediate wall crossings cancel out. The point is, the intermediate steps
are completely irrelevant except for the single purpose of mediating a transition
from $\Theta_{\rm emit}$ to $\Theta_{\rm obs}$. A sequence of intermediate states 
must exist to realize the initial and final vacuum values. Conversely, 
if $\Theta_{\rm emit}$ and $\Theta_{\rm obs}$ correspond to vacua of the theory, 
then there exists a sequence of transitions connecting them.
Further, when there is finitely many possible values, which 
support fast bubble nucleations with $\Gamma\sim H_0^4$ as proposed in
\cite{Kaloper:2025goq,Kaloper:2026slg,Kaloper:2026gib} the initial value
$\Theta_{\rm emit}$ will naturally evolve to the value $\Theta_{\rm obs}$ at a later time,
completing the sequence.

Therefore the observable
rotation for light coming from any direction 
depends only on the vacuum at emission and the vacuum at
observation. It is completely independent of the number of crossings,
the geometry of the wall network, the ordering of intermediate vacua,
and the detailed history of the walls. 
Hence equation \eqref{endpointrotation} immediately implies that
vacuum-interface birefringence is not a random-walk phenomenon.
Additional wall crossings do not increase the magnitude of the signal.
Likewise the effect does not accumulate with propagation distance.
The observable is controlled by endpoint data rather than by the path
connecting them. 

Note also that an observable signal does not require a domain wall to be present 
today. It is sufficient that the vacuum at emission differ from the vacuum 
at observation. Once a photon has crossed the relevant interfaces, the 
resulting phase shift is permanently imprinted in its polarization state. 
The walls responsible for the transition may subsequently disappear without erasing the signal.
Thus the observable birefringence arises because the vacuum on the last 
scattering surface differs from the vacuum today. 
The observations therefore probe vacuum history rather than the current
abundance of domain walls.

These results acquire a particularly simple cosmological interpretation in
the class of scenarios which motivated this work. In our case, first of all,
in the absence of walls $\Theta$ is a fixed, rigid parameter. When we further take that
\bi
\item the entire observable universe originated from a single inflationary patch, with
a single vacuum;

\item the vacuum structure of the low-energy theory is labelled by a
vacuum parameter $\Theta$;

\item prior to a dark-sector phase transition the vacua were degenerate, as is the case
in generic QCD-like theories;

\item either no axion exists in the relevant low-energy theory or any
axion associated with the vacuum structure is sufficiently heavy that it
can be integrated out and so does not fluctuate.
\ei
We further suppose that vacuum transitions begin only after last
scattering and terminate before the present epoch, if they are to 
affect the CMB. Then all CMB photons
originate in the same vacuum on the last scattering surface, which we
denote by $\Theta_{\rm LSS}$, while all observers today reside in a
common final vacuum $\Theta_0$ . In this case Eq.~\eqref{endpointrotation} reduces to
\be
\Delta\vartheta(\hat n)
=
\frac{\zeta}{6{\cal M}^2}
\Bigl(
\Theta_0-\Theta_{\rm LSS}
\Bigr)
\label{uniformrotation}
\ee
for every direction $\hat n$ on the sky. The generic prediction is 
therefore an {\it isotropic birefringence
signal}. The leading observable is a pure monopole rotation of the CMB
polarization pattern. 
Although many vacuum interfaces may have existed
between last scattering and the present epoch, their individual
contributions cancel, leaving only the mismatch between the initial and
final vacua.

In the simplest two-vacuum realization which we focused on for the most 
part here and in \cite{Kaloper:2025goq,Kaloper:2026slg,Kaloper:2026gib},
$\Theta_{\rm obs}-\Theta_{\rm emit}$ is the same for every line of sight.
Hence the net crossing number is identical in all directions and the
resulting birefringence is spatially uniform. The same conclusion holds if there are multiple vacua.
The detailed arrangement of intermediate walls is irrelevant. 

Indeed, this follows from an even simpler topological argument, simplifying the argument above for the
case when there are only two vacua, and the phase transition rate is fast allowing the presence of 
many bubbles along the way. Since the initial vacuum is the same in any direction, 
and there are no axionic cosmic strings because axion is too heavy, 
the total result in acquisition of phase is independent of path and only depends on 
end points. The formal way to state this is that the Berry holonomy vanishes identically, as discussed earlier.
But not the Pancharatnam phase, which accumulates from interface crossings  with one sign coming into the
bubble and the other sign coming out. Since the net crossing number, 
$N_+ - N_-$, must equal the vacuum mismatch between emission and observation, 
and since in the present two-vacuum realization this mismatch is unity, we have 
\be N_+ - N_- = 1 \, . \ee 
the total accumulated phase is exactly the same along
every line of sight.

This prediction differs qualitatively from scenarios involving
additional light dynamical degrees of freedom. Ultralight axions,
axionic field fluctuations, string-wall systems, incomplete inflationary
preparation, vacuum structures intersecting the last scattering surface,
or other sources of nontrivial topology may all generate anisotropic
birefringence and nontrivial angular power spectra. Such effects arise
from additional dynamical ingredients beyond the minimal interface
mechanism considered here.

Within the framework studied in this work, the leading prediction is not
a microscopically induced random rotation field generated by accumulated wall crossings, but
a uniform polarization rotation determined solely by the vacuum mismatch
between emission and observation. A more detailed study of the precise implications for 
the observables to be probed by future searches is currently under investigation and will be presented
elsewhere.

\section{Summary}

Recent claims of weak optical activity in the cosmic microwave background
\cite{Komatsu:2022nvu}, at the level
\be 
\Delta \vartheta \sim 10^{-3}\,{\rm radians} \, ,
\label{variations}
\ee
motivate a careful reassessment of possible sources of cosmic birefringence. While
common approaches to this phenomenon invoke ultralight fields that vary slowly
over cosmological times, other attractive explanations are viable.
Effects of this magnitude can naturally arise from electromagnetic interactions
with vacuum interfaces in the dark sector, without the need for ultralight fields
beyond electromagnetism itself. A recently proposed example involves evanescent
dark-energy domain walls carrying a Chern--Simons coupling of order
$\zeta{\cal Q}/(6{\cal M}^2)\sim{\rm few}\times10^{-3}$
\cite{Kaloper:2025goq,Kaloper:2026slg,Kaloper:2026gib}. 
The magnitude of this coupling can be naturally of the order of the
anomaly-induced Chern--Simons couplings, $\sim \alpha_{\rm QED}/2\pi$ when there are kinetic 
top-form mixings between the visible and dark sectors, and 
there are heavy charged states which are integrated out. 
Therefore our approach does not require introducing ultralight scales 
beyond those already present in the dark sector. 

To elucidate the physics of this phenomenon, in this work we developed a unified
quantum-mechanical description of photon propagation that applies both to thick
axion domain walls and to sharp interfaces generated by membrane-induced jumps of
a vacuum angle. In all cases, the essential criterion for polarization rotation
is whether a photon crosses a boundary separating distinct vacua. Photons
propagating entirely within a single vacuum region experience no rotation. When
a boundary is crossed, the magnitude of the effect depends only on the discrete
change of the vacuum parameter across the interface and is independent of the
detailed wall profile.

Formulating photon propagation as a one-dimensional scattering problem makes the
origin of this behavior transparent. The wall-localized Chern--Simons interaction
introduces a factor of the photon frequency, while the outgoing GreenÕs function
contributes a compensating inverse power. Their cancellation ensures that the
net effect of the interface is a pure phase shift, independent of photon
wavelength below a physical ultraviolet cutoff. This provides a dynamical
explanation for the frequency-independent polarization rotation found in the
effective theory.

The resulting polarization rotation is completely fixed by the normalization of
the wall Chern--Simons operator and is therefore protected by an emergent
higher-form symmetry of the low-energy effective theory. As a consequence, the
effect is localized at vacuum interfaces, robust against smooth deformations of
the wall, and insensitive to microscopic details as long as the effective
description remains valid. For extremely thin domain walls, the rotation does
not accumulate continuously with distance or redshift, but instead appears
discretely when vacuum transitions are encountered. 

This sharply distinguishes the present mechanism from conventional axion
birefringence. In axion electrodynamics, polarization rotation is a dynamical
effect controlled by a broken $0$-form symmetry and depends on the spacetime
evolution of a light scalar, requiring ultralight axions with masses around or below about 
$10^{-28}\,{\rm eV}$ and a detailed cosmological history. By contrast, the effect
described here persists even when all light axionic degrees of freedom are absent
and does not rely on background time evolution. 

Several further aspects of this mechanism merit emphasis. Because the polarization
rotation is induced by vacuum interfaces rather than accumulated dynamically
over cosmological timescales, it does not require ultralight axions to be operative
throughout cosmic history. The adiabatic regime for phase rotation requires only
that the photon frequency exceed the axion mass. Consequently, adiabatic
transport of CMB polarization is compatible with axion masses far above the
Hubble scale today, provided they remain below CMB frequencies. Even an axion
with a mass as large as $m_\phi \sim 10^{-4}\,{\rm eV}$ could adiabatically rotate
CMB polarization if supported by a domain wall, a regime often overlooked in
astrophysical discussions. The walls which can do this would be thick compared to
the ultrathin walls which induce the Pancharatman phase, at last about a millimeter 
across. Nonetheless, such solitonic axion configurations
would still appear extremely thin compared to the vast regions typically
associated with ultralight fields.

These considerations highlight that the key physical requirement for polarization
rotation is not the long-term persistence of light degrees of freedom, but the
existence of vacuum interfaces encountered by CMB photons along their line of
sight. Once the effect is understood as induced by, or literally localized, at such interfaces, 
questions about the subsequent cosmological evolution of the walls become largely
decoupled from the polarization imprint itself. While the existence of domain
walls must of course be consistent with other cosmological tests and bounds, such as
whether they can be produced, survive till late times and so on, 
such considerations constrain the viability of specific wall scenarios rather
than the mechanism of polarization rotation.

Moreover, the domain walls responsible for the polarization twist need not
persist to the present epoch. If vacuum interfaces nucleated copiously after the
last scattering surface and subsequently disappeared, as is natural in some
early dark energy-like scenarios \cite{Kaloper:2025goq}, the resulting polarization rotation would
nevertheless be imprinted on the CMB from every direction of the sky. To reach
observers residing in the final vacuum, all CMB photons must have crossed at
least some walls along their trajectory. Walls which are present but decay prior to 
last scattering, or grow out of the horizon 
are irrelevant for this signal, since the CMB photons did not originate before
that epoch.

The symmetry structure of the infrared theory is most consistently understood 
not as a collection of unrelated symmetries that appear or disappear across 
scales, but as a single intertwined symmetry framework whose realization 
depends on the light field content. In the ultraviolet completion, the theory 
possesses the usual gauge symmetries of electromagnetism, the shift symmetry 
of the axion sector, and the emergent gauge invariance 
of the dual $3$-form fields associated 
with the dark sector. After integrating out heavy degrees of freedom, 
these symmetries are not lost or independently generated in the infrared; 
instead, they become linked in a way that ties their transformations 
together through the electromagnetic topological structure.

In this picture, the electromagnetic $1$-form symmetry, the axion shift symmetry, 
and the dual $3$-form gauge symmetries are not independent in the low-energy theory. 
Rather, they are different aspects of a single combined symmetry structure in which 
transformations in one sector necessarily induce correlated transformations in the others. 
The key ingredient that enforces this linkage is the electromagnetic topological density, 
encoded through the relation $d\Upsilon^{(3)} = \frac{1}{8\pi^2} F \wedge F$. 
This relation fixes how electromagnetic topology enters the consistency 
conditions of the low-energy theory and prevents the symmetry 
transformations from factorizing into independent pieces.

From this perspective, the appropriate description is that the theory possesses 
a single underlying symmetry structure whose different aspects become more or 
less explicit depending on which degrees of freedom are retained in the effective 
description. At the level of fields, this structure is realized through modified gauge 
transformations in which shifts of the axion field and the $3$-form potentials must 
be accompanied by compensating contributions proportional to the electromagnetic 
topological sector. The resulting infrared theory therefore does not admit a clean 
separation between the different symmetry sectors: electromagnetic topology 
acts as the organizing constraint that ties them together 
into a single consistent framework.

While cosmology has long searched for signatures of nontrivial topology of
spacetime itself, the results presented here suggest that observations may also
be sensitive to the topology of ``inner space'' encoded in the vacuum structure
of the dark sector. It is intriguing that a phenomenon of interface-localized polarization rotation 
familiar from optical activity in condensed matter systems may have a counterpart in the dark sector
of the universe, potentially imprinted on the sky and accessible to future
polarization measurements.

\vskip1cm

{\bf Acknowledgments}: 
This research was supported in part by the DOE Grant DE-SC0009999.


\begin{thebibliography}{99}

%\cite{Komatsu:2022nvu}
\bibitem{Komatsu:2022nvu}
E.~Komatsu,
``New physics from the polarized light of the cosmic microwave background,''
Nature Rev. Phys. \textbf{4}, no.7, 452-469 (2022)
%doi:10.1038/s42254-022-00452-4
[arXiv:2202.13919 [astro-ph.CO]].
%179 citations counted in INSPIRE as of 18 Jan 2026

%\cite{Huang:1985tt}
\bibitem{Huang:1985tt}
M.~C.~Huang and P.~Sikivie,
``The Structure of Axionic Domain Walls,''
Phys. Rev. D \textbf{32}, 1560 (1985). 
%doi:10.1103/PhysRevD.32.1560
%120 citations counted in INSPIRE as of 27 Jan 2026

%\cite{Harari:1992ea}
\bibitem{Harari:1992ea}
D.~Harari and P.~Sikivie,
``Effects of a Nambu-Goldstone boson on the polarization 
of radio galaxies and the cosmic microwave background,''
Phys. Lett. B \textbf{289}, 67-72 (1992). 
%doi:10.1016/0370-2693(92)91363-E
%300 citations counted in INSPIRE as of 27 Jan 2026

%\cite{Carroll:1998zi}
\bibitem{Carroll:1998zi}
S.~M.~Carroll,
``Quintessence and the rest of the world,''
Phys. Rev. Lett. \textbf{81}, 3067-3070 (1998)
%doi:10.1103/PhysRevLett.81.3067
[arXiv:astro-ph/9806099 [astro-ph]].
%1291 citations counted in INSPIRE as of 02 May 2026

%\cite{Lue:1998mq}
\bibitem{Lue:1998mq}
A.~Lue, L.~M.~Wang and M.~Kamionkowski,
``Cosmological signature of new parity violating interactions,''
Phys. Rev. Lett. \textbf{83}, 1506-1509 (1999)
%doi:10.1103/PhysRevLett.83.1506
[arXiv:astro-ph/9812088 [astro-ph]].
%758 citations counted in INSPIRE as of 09 Apr 2026

%\cite{Pospelov:2008gg}
\bibitem{Pospelov:2008gg}
M.~Pospelov, A.~Ritz and C.~Skordis,
``Pseudoscalar perturbations and polarization of the cosmic microwave background,''
Phys. Rev. Lett. \textbf{103}, 051302 (2009)
%doi:10.1103/PhysRevLett.103.051302
[arXiv:0808.0673 [astro-ph]].
%130 citations counted in INSPIRE as of 09 Apr 2026

%\cite{Minami:2020odp}
\bibitem{Minami:2020odp}
Y.~Minami and E.~Komatsu,
``New Extraction of the Cosmic Birefringence from the Planck 2018 Polarization Data,''
Phys. Rev. Lett. \textbf{125}, no.22, 221301 (2020)
%doi:10.1103/PhysRevLett.125.221301
[arXiv:2011.11254 [astro-ph.CO]].
%312 citations counted in INSPIRE as of 02 May 2026

%\cite{Takahashi:2020tqv}
\bibitem{Takahashi:2020tqv}
F.~Takahashi and W.~Yin,
``Kilobyte Cosmic Birefringence from ALP Domain Walls,''
JCAP \textbf{04}, 007 (2021)
%doi:10.1088/1475-7516/2021/04/007
[arXiv:2012.11576 [hep-ph]].
%87 citations counted in INSPIRE as of 09 Apr 2026

%\cite{Diego-Palazuelos:2022dsq}
\bibitem{Diego-Palazuelos:2022dsq}
P.~Diego-Palazuelos, J.~R.~Eskilt, Y.~Minami, M.~Tristram, R.~M.~Sullivan,
A.~J.~Banday, R.~B.~Barreiro, H.~K.~Eriksen, K.~M.~G{\'o}rski and R.~Keskitalo, \textit{et al.}
``Cosmic Birefringence from the Planck Data Release 4,''
Phys. Rev. Lett. \textbf{128}, no.9, 091302 (2022)
%doi:10.1103/PhysRevLett.128.091302
[arXiv:2201.07682 [astro-ph.CO]].
%178 citations counted in INSPIRE as of 02 May 2026

%\cite{Ferreira:2023jbu}
\bibitem{Ferreira:2023jbu}
R.~Z.~Ferreira, S.~Gasparotto, T.~Hiramatsu, I.~Obata and O.~Pujolas,
``Axionic defects in the CMB: birefringence and gravitational waves,''
JCAP \textbf{05}, 066 (2024)
%doi:10.1088/1475-7516/2024/05/066
[arXiv:2312.14104 [hep-ph]].
%44 citations counted in INSPIRE as of 01 Apr 2026

%\cite{Pancharatnam:1956url}
\bibitem{Pancharatnam:1956url}
S.~Pancharatnam,
``Generalized theory of interference, and its applications,''
Proc. Indian Acad. Sci. A \textbf{44}, no.5, 247-262 (1956).
%doi:10.1007/BF03046050
%272 citations counted in INSPIRE as of 07 Apr 2026

%\cite{Pancharatnam:1956url2}
\bibitem{Pancharatnam:1956url2}
S. Pancharatnam, 
``Generalized theory of interference and its applications. 
part ii. partially coherent pencils,'' 
Proc. Indian Acad. Sci. - Section A \textbf{44}, 398Ð417 (1956).

%\cite{Kaloper:2025goq}
\bibitem{Kaloper:2025goq}
N.~Kaloper,
``Discretely evanescent dark energy,''
JCAP \textbf{11}, 075 (2025) 
%doi:10.1088/1475-7516/2025/11/075
[arXiv:2506.04317 [hep-th]].
%2 citations counted in INSPIRE as of 09 Jan 2026

%\cite{Kaloper:2026slg}
\bibitem{Kaloper:2026slg}
N.~Kaloper,
``Electromagnetic Couplings of Dark Domain Walls,''
[arXiv:2602.03933 [hep-th]].
%1 citations counted in INSPIRE as of 09 Apr 2026

%\cite{Kaloper:2026gib}
\bibitem{Kaloper:2026gib}
N.~Kaloper,
``Cloaking Cosmic Light,''
[arXiv:2602.20248 [astro-ph.CO]].
%0 citations counted in INSPIRE as of 09 Apr 2026

%\cite{Adler:1969er}
\bibitem{Adler:1969er}
S.~L.~Adler and W.~A.~Bardeen,
``Absence of higher order corrections in the anomalous axial vector divergence equation,''
Phys. Rev. \textbf{182}, 1517-1536 (1969). 
%doi:10.1103/PhysRev.182.1517
%1327 citations counted in INSPIRE as of 14 Jul 2026
Copy to ClipboardDownload


%\cite{Ganoulis:1986rd}
\bibitem{Ganoulis:1986rd}
N.~Ganoulis and M.~Hatzis,
``Light Scattering by an Axionic Domain Wall,''
Mod. Phys. Lett. A \textbf{1}, 409 (1986). 
%doi:10.1142/S0217732386000518
%4 citations counted in INSPIRE as of 31 May 2026

%\cite{Favitta:2023hlx}
\bibitem{Favitta:2023hlx}
A.~M.~Favitta, I.~H.~Brevik and M.~M.~Chaichian,
``Axion electrodynamics: Green{\textquoteright}s functions, zero-point energy and optical activity,''
Annals Phys. \textbf{455}, 169396 (2023)
%doi:10.1016/j.aop.2023.169396
[arXiv:2302.13129 [hep-th]].
%14 citations counted in INSPIRE as of 31 May 2026

%\cite{Agrawal:2023sbp}
\bibitem{Agrawal:2023sbp}
P.~Agrawal and A.~Platschorre,
``The monodromic axion-photon coupling,''
JHEP \textbf{01}, 169 (2024)
%doi:10.1007/JHEP01(2024)169
[arXiv:2309.03934 [hep-th]].
%24 citations counted in INSPIRE as of 19 Apr 2026

%\cite{Blasi:2024xvj}
\bibitem{Blasi:2024xvj}
S.~Blasi,
``Monodromic transparency of axion domain walls,''
JHEP \textbf{05}, 013 (2025)
%doi:10.1007/JHEP05(2025)013
[arXiv:2412.15085 [hep-ph]].
%4 citations counted in INSPIRE as of 19 Apr 2026

%\cite{Gaiotto:2014kfa}
\bibitem{Gaiotto:2014kfa}
D.~Gaiotto, A.~Kapustin, N.~Seiberg and B.~Willett,
``Generalized Global Symmetries,''
JHEP \textbf{02}, 172 (2015)
%doi:10.1007/JHEP02(2015)172
[arXiv:1412.5148 [hep-th]].
%1719 citations counted in INSPIRE as of 09 Apr 2026

%\cite{Hidaka:2020iaz}
\bibitem{Hidaka:2020iaz}
Y.~Hidaka, M.~Nitta and R.~Yokokura,
``Higher-form symmetries and 3-group in axion electrodynamics,''
Phys. Lett. B \textbf{808}, 135672 (2020)
%doi:10.1016/j.physletb.2020.135672
[arXiv:2006.12532 [hep-th]].
%75 citations counted in INSPIRE as of 19 Apr 2026

%\cite{Brennan:2020ehu}
\bibitem{Brennan:2020ehu}
T.~D.~Brennan and C.~Cordova,
``Axions, higher-groups, and emergent symmetry,''
JHEP \textbf{02}, 145 (2022)
%doi:10.1007/JHEP02(2022)145
[arXiv:2011.09600 [hep-th]].
%100 citations counted in INSPIRE as of 19 Apr 2026

\bibitem{messiah}
A. Messiah, {\it Quantum Mechanics vols. I $\&$ II}, Dover, Mineola, NY 1999. 

%\cite{Kaloper:2025wgn}
\bibitem{Kaloper:2025wgn}
N.~Kaloper,
``Alternative to axions,''
Phys. Rev. D \textbf{113}, no.1, L011701 (2026)
%doi:10.1103/4k8p-gpgr
[arXiv:2504.21078 [hep-ph]].
%6 citations counted in INSPIRE as of 02 Feb 2026

%\cite{Kaloper:2025upu}
\bibitem{Kaloper:2025upu}
N.~Kaloper, 
``A Quantal Theory of Restoration of Strong CP Symmetry,''
[arXiv:2505.04690 [hep-ph]].
%1 citations counted in INSPIRE as of 25 May 2025

%\cite{Kaloper:2008qs}
\bibitem{Kaloper:2008qs}
N.~Kaloper and L.~Sorbo,
``Where in the String Landscape is Quintessence,''
Phys. Rev. D \textbf{79}, 043528 (2009)
%doi:10.1103/PhysRevD.79.043528
[arXiv:0810.5346 [hep-th]].
%94 citations counted in INSPIRE as of 25 May 2025

%\cite{Kaloper:2008fb}
\bibitem{Kaloper:2008fb}
N.~Kaloper and L.~Sorbo,
``A Natural Framework for Chaotic Inflation,''
Phys. Rev. Lett. \textbf{102}, 121301 (2009)
%doi:10.1103/PhysRevLett.102.121301
[arXiv:0811.1989 [hep-th]].
%398 citations counted in INSPIRE as of 09 Jan 2026

%\cite{Kaloper:2011jz}
\bibitem{Kaloper:2011jz}
N.~Kaloper, A.~Lawrence and L.~Sorbo,
``An Ignoble Approach to Large Field Inflation,''
JCAP \textbf{03}, 023 (2011)
%doi:10.1088/1475-7516/2011/03/023
[arXiv:1101.0026 [hep-th]].
%310 citations counted in INSPIRE as of 09 Jan 2026

%\cite{Luscher:1978rn}
\bibitem{Luscher:1978rn}
M.~L\"uscher, ``The Secret Long Range Force in Quantum Field Theories With Instantons,''
Phys. Lett. B \textbf{78}, 465-467 (1978).
%doi:10.1016/0370-2693(78)90487-2
%158 citations counted in INSPIRE as of 12 Aug 2024
 
%\cite{Gabadadze:1997kj}
\bibitem{Gabadadze:1997kj}
G.~Gabadadze,
``Modeling the glueball spectrum by a closed bosonic membrane,''
Phys. Rev. D \textbf{58}, 094015 (1998)
%doi:10.1103/PhysRevD.58.094015
[arXiv:hep-ph/9710402 [hep-ph]].
%34 citations counted in INSPIRE as of 20 Aug 2024

%\cite{Gabadadze:2002ff}
\bibitem{Gabadadze:2002ff}
G.~Gabadadze and M.~Shifman,
``QCD vacuum and axions: What's happening?,''
Int. J. Mod. Phys. A \textbf{17}, 3689-3728 (2002)
%doi:10.1142/S0217751X02011357
[arXiv:hep-ph/0206123 [hep-ph]].
%70 citations counted in INSPIRE as of 20 Aug 2024

%\cite{Dvali:2005an}
\bibitem{Dvali:2005an}
G.~Dvali, ``Three-form gauging of axion symmetries and gravity,''
[arXiv:hep-th/0507215 [hep-th]].
%143 citations counted in INSPIRE as of 19 Apr 2023

%\cite{Dvali:2005zk}
\bibitem{Dvali:2005zk}
G.~Dvali,
``A Vacuum accumulation solution to the strong CP problem,''
Phys. Rev. D \textbf{74}, 025019 (2006)
%doi:10.1103/PhysRevD.74.025019
[arXiv:hep-th/0510053 [hep-th]].
%56 citations counted in INSPIRE as of 21 Jan 2026

%\cite{vonDossow:2025bwr}
\bibitem{vonDossow:2025bwr}
R.~M.~von Dossow, A.~Mart{\'\i}n-Ruiz and L.~F.~Urrutia,
``Higher-Order Derivative Corrections to Axion Electrodynamics in 3D Topological Insulators,''
Symmetry \textbf{17}, no.4, 581 (2025). 
%doi:10.3390/sym17040581
%4 citations counted in INSPIRE as of 19 Apr 2026

%\cite{Vilenkin:2000jqa}
\bibitem{Vilenkin:2000jqa}
A.~Vilenkin and E.~P.~S.~Shellard,
{\it Cosmic Strings and Other Topological Defects}, 
Cambridge University Press, Cambridge, UK, 2000. 
%ISBN 978-0-521-65476-0
%107 citations counted in INSPIRE as of 12 Apr 2026

%\cite{Peccei:1977hh}
\bibitem{Peccei:1977hh}
R.~D.~Peccei and H.~R.~Quinn,
``CP Conservation in the Presence of Instantons,''
Phys. Rev. Lett. \textbf{38}, 1440-1443 (1977).
%doi:10.1103/PhysRevLett.38.1440
%7977 citations counted in INSPIRE as of 13 Aug 2024

%\cite{Weinberg:1977ma}
\bibitem{Weinberg:1977ma}
S.~Weinberg,
``A New Light Boson?,''
Phys. Rev. Lett. \textbf{40}, 223-226 (1978).
%doi:10.1103/PhysRevLett.40.223
%5665 citations counted in INSPIRE as of 13 Aug 2024

%\cite{Wilczek:1977pj}
\bibitem{Wilczek:1977pj}
F.~Wilczek,
``Problem of Strong  $P$  and  $T$  Invariance in the Presence of Instantons,''
Phys. Rev. Lett. \textbf{40}, 279-282 (1978).
%doi:10.1103/PhysRevLett.40.279
%5431 citations counted in INSPIRE as of 13 Aug 2024

%\cite{Wilczek:1987mv}
\bibitem{Wilczek:1987mv}
F.~Wilczek,
``Two Applications of Axion Electrodynamics,''
Phys. Rev. Lett. \textbf{58}, 1799 (1987).
%doi:10.1103/PhysRevLett.58.1799
%650 citations counted in INSPIRE as of 10 Apr 2026

%\cite{Raffelt:1987im}
\bibitem{Raffelt:1987im}
G.~Raffelt and L.~Stodolsky,
``Mixing of the Photon with Low Mass Particles,''
Phys. Rev. D \textbf{37}, 1237 (1988). 
%doi:10.1103/PhysRevD.37.1237
%1094 citations counted in INSPIRE as of 10 Apr 2026

%\cite{Simon:1983mh}
\bibitem{Simon:1983mh}
B.~Simon,
``Holonomy, the quantum adiabatic theorem, and Berry's phase,''
Phys. Rev. Lett. \textbf{51}, 2167-2170 (1983).
%doi:10.1103/PhysRevLett.51.2167
%933 citations counted in INSPIRE as of 07 Apr 2026

%\cite{Berry:1984jv}
\bibitem{Berry:1984jv}
M.~V.~Berry,
``Quantal phase factors accompanying adiabatic changes,''
Proc. Roy. Soc. Lond. A \textbf{392}, 45-57 (1984).
%doi:10.1098/rspa.1984.0023
%2482 citations counted in INSPIRE as of 07 Apr 2026

%\cite{Kaloper:2023kua} 
\bibitem{Kaloper:2023kua}
N.~Kaloper,
``Axion flux monodromy discharges relax the cosmological constant,''
JCAP \textbf{11}, 032 (2023)
%doi:10.1088/1475-7516/2023/11/032
[arXiv:2307.10365 [hep-th]].
%3 citations counted in INSPIRE as of 12 Aug 2024

%\cite{Born:1999ory}
\bibitem{Born:1999ory}
M.~Born and E.~Wolf,
{\it Principles of Optics}, 7th ed., Cambridge University Press, Cambridge, UK 1999.
%ISBN 978-1-139-64418-1
%doi:10.1017/CBO9781139644181
%89 citations counted in INSPIRE as of 05 May 2026

%\cite{Amendola:2005ad} 
\bibitem{Amendola:2005ad}
L.~Amendola and R.~Barbieri,
``Dark matter from an ultra-light pseudo-Goldsone-boson,''
Phys. Lett. B \textbf{642}, 192-196 (2006)
%doi:10.1016/j.physletb.2006.08.069
[arXiv:hep-ph/0509257 [hep-ph]].
%224 citations counted in INSPIRE as of 04 May 2026

%\cite{Hlozek:2014lca}
\bibitem{Hlozek:2014lca}
R.~Hlozek, D.~Grin, D.~J.~E.~Marsh and P.~G.~Ferreira,
``A search for ultralight axions using precision cosmological data,''
Phys. Rev. D \textbf{91}, no.10, 103512 (2015)
%doi:10.1103/PhysRevD.91.103512
[arXiv:1410.2896 [astro-ph.CO]].
%531 citations counted in INSPIRE as of 04 May 2026

\end{thebibliography}
\end{document}